\documentclass[useAMS,twocolumn,usenatbib]{mnras}
\pdfoutput=1
\usepackage{times,psfig,epsfig}
\usepackage{rotating, graphicx}
\usepackage{color}
\usepackage{amssymb, amsmath}
\usepackage[normalem]{ulem}
\usepackage[titletoc]{appendix}
\usepackage{adjustbox}
\usepackage{booktabs}
\def\aap{A\&A}

\def\apjl{ApJL}

\def\apjs{ApJS}

\newcommand{\bdm}{\begin{displaymath}}
\newcommand{\edm}{\end{displaymath}}
\newcommand{\beq}{\begin{equation}}
\newcommand{\eeq}{\end{equation}}
\newcommand{\beqnarr}{\begin{eqnarray}}
\newcommand{\eeqnarr}{\end{eqnarray}}
\newcommand{\bit}{\begin{itemize}}
\newcommand{\eit}{\end{itemize}}
\newcommand{\ben}{\begin{enumerate}}
\newcommand{\een}{\end{enumerate}}
\newcommand{\bfi}{\begin{figure}[htb]}
\newcommand{\bpfi}{\begin{figure}[p]}
\newcommand{\barr}{\begin{array}}
\newcommand{\earr}{\end{array}}
\newcommand{\bec}{\begin{center}}
\newcommand{\eec}{\end{center}}
\newcommand{\bs}{\begin{sideways}}
\newcommand{\es}{\end{sideways}}

 \newcommand{\mincir}{\raise
  -2.truept\hbox{\rlap{\hbox{$\sim$}}\raise5.truept \hbox{$<$}\ }}
\newcommand{\magcir}{\raise
  -2.truept\hbox{\rlap{\hbox{$\sim$}}\raise5.truept \hbox{$>$}\ }}
\newcommand{\siml}{\raise
  -2.truept\hbox{\rlap{\hbox{$\sim$}}\raise5.truept \hbox{$<$}\ }}
\newcommand{\simg}{\raise
  -2.truept\hbox{\rlap{\hbox{$\sim$}}\raise5.truept \hbox{$>$}\ }}

\title[SMBH -- hot atmosphere correlations in TNG]{Correlations between supermassive black holes and hot gas atmospheres in IllustrisTNG and X-ray observations} %I'm proposing this title NW
\author[N. Truong et al]{Nhut Truong$^{1,2}$\thanks{truong@mpia-hd.mpg.de}, Annalisa Pillepich$^1$, and Norbert Werner$^{3,4,2}$
 \\~\\
\footnotesize
$^1$ Max-Planck-Institut f\"ur Astronomie, K\"onigstuhl 17, D-69117 Heidelberg, Germany\\
$^2$ MTA-E\"otv\"os University Lend\"ulet Hot Universe Research Group, P\'azm\'any P\'eter s\'et\'any 1/A, Budapest, 1117, Hungary \\
$^3$Department of Theoretical Physics and Astrophysics, Faculty of Science, Masaryk University, Kotl\'a\v{r}sk\'a 2, Brno, 611 37, Czech Republic \\
$^4$School of Science, Hiroshima University, 1-3-1 Kagamiyama, Higashi-Hiroshima 739-8526, Japan \\
}

\begin{document}
\maketitle

%%%%%%%%%%%%%%%
% Abstract    %
%%%%%%%%%%%%%%%
\begin{abstract}
Recent X-ray observations have revealed remarkable correlations between the masses of central supermassive black holes (SMBHs) and the X-ray properties of the hot atmospheres permeating their host galaxies, thereby indicating the crucial role of the atmospheric gas in tracing SMBH growth in the high-mass regime. We examine this topic theoretically using the IllustrisTNG cosmological simulations and provide insights to the nature of this SMBH -- gaseous halo connection. By carrying out a mock X-ray analysis for a mass-selected sample of TNG100 simulated galaxies at $z=0$, we inspect the relationship between the masses of SMBHs and the hot gas temperatures and luminosities at various spatial and halo scales -- from galactic ($\sim R_{\rm e}$) to group/cluster scales ($\sim R_{\rm 500c}$). We find strong SMBH-X-ray correlations mostly in quenched galaxies and find that the correlations become stronger and tighter at larger radii. Critically, the X-ray temperature ($k_{\rm B}T_{\rm X}$) at large radii ($r\gtrsim5R_{\rm e}$) traces the SMBH mass with a remarkably small scatter ($\sim0.2$ dex). The relations emerging from IllustrisTNG are broadly consistent with those obtained from recent X-ray observations. Overall, our analysis suggests that, within the framework of IllustrisTNG, the present-time $M_{\rm BH}-k_{\rm B}T_{X}$ correlations at the high-mass end ($M_{\rm BH}\gtrsim 10^8 M_\odot$) are fundamentally a reflection of the SMBH mass -- halo mass relation, which at such high masses is set by the hierarchical assembly of structures. The exact form, locus, and scatter of those scaling relations are, however, sensitive to feedback processes such as those driven by star formation and SMBH activity.
\end{abstract}
%%%%%%%%%%%%%%%%%%%%%%%%%%%%%%%%%%%%%%%%%%%%%%%%%%%%%%%%%%%%%%%%%%%
\section{Introduction}
\label{sec:intro}
%%%%%%%%%%%%%%%%%%%%%%%%%%%%%%%%%%%%%%%%%%%%%%%%%%%%%%%%%%%%%%%%%%%
Supermassive black holes (SMBHs) are nearly ubiquitous in massive galaxies in the local Universe (see \citealt{kormendy.ho.2013,saglia.etal.2016,van.den.bosch.2016} for recent SMBH catalogues). They are thought to be the central engines powering the feedback from active galactic nuclei (AGN), which is a standard ingredient in cosmological galaxy simulations for its crucial role in quenching star formation as well as partially shaping the thermodynamics of the hot atmospheres permeating massive galaxies as well as groups and clusters of galaxies (\citealt{springel.etal.2005,booth.schaye.2009,McCarthy.etal.2010,planelles.etal.2014,schaye.etal.2015,choi.etal.2015,weinberger.etal.2017, henden.etal.2018,dave.etal.2019}). Understanding the interaction between SMBHs and their host galaxies, as well as their overall co-evolution, is thus particularly important for our understanding of galaxy formation and evolution. This can be done, albeit indirectly, by inspecting the correlations between SMBH masses and galaxy properties (see \citealt{kormendy.ho.2013} for a review).

It has been established observationally that SMBHs strongly correlate with the properties of the stellar components of their host galaxies (bulge mass, bulge luminosity, and stellar velocity dispersion; \citealt{magorrian.etal.1998,gebhardt.etal.2000,haring.rix.2004,gultekin.etal.2009,mcconnell.ma.2013}). The existence of these correlations suggests the existence of a close interaction between the AGN and the interstellar medium in the central region of the host galaxy (\citealt{silk.rees.1998,king.2003,churazov.etal.2005}). They also serve as valuable means of estimating the SMBHs mass via the measurements of their observable stellar proxies. However, at the high-mass end, it is intriguing that the SMBHs appear to be over massive with respect to the predictions of their stellar mass proxies (\citealt{gebhardt.etal.2011,mcconnell.etal.2011,hlavacek-larrondo.etal.2012,ferre-mateu.etal.2015,savorgnan.etal.2015}).

Recently, X-ray observations have revealed remarkable correlations between the central SMBH masses and the X-ray properties of the hot atmospheres at various spatial and galaxy/halo scales (\citealt{bogdan.etal.2018,phipps.etal.2019,lakhchaura.etal.2019,gaspari.etal.2019,martin-navarro.etal.2020}). Importantly, these studies show that at the high-mass end (SMBH masses $\gtrsim 5-10\times 10^7 M_\odot$), the X-ray quantities, in particular the X-ray gas temperature, correlate with the SMBH mass even better than the bulge mass and stellar velocity dispersion. These results suggest that the hot atmospheric gas could be the best tracer of SMBH growth at the high-mass end. 

At present, there is no general consensus regarding the explanation for the origin of the correlations between the SMBH mass and the X-ray properties of the hot atmospheres. \cite{bogdan.etal.2018} interpret the correlation between the central SMBH mass and the intra-cluster medium (ICM) temperature ($k_{\rm B}T_{\rm 500c}$) as a result of cluster-scale processes such as accreted cold gas inflows or mergers that feed the SMBH growth. \cite{lakhchaura.etal.2019} only found significant correlations for brightest group and cluster galaxies (BGGs/BCGs) or cored galaxies, as opposed to non-BCGs and lenticular (S0) galaxies, making them speculate that the correlations might be established primarily by mergers. This line of interpretation favors a non-causal picture of SMBH-galaxy co-evolution, in which the SMBH and its host galaxy grow simultaneously by mergers, and the observed correlations arise naturally as a consequence of the central-limit theorem (\citealt{peng.2007,jahnke.maccio.2011,volonteri.etal.2011}). 

On the other hand, \cite{gaspari.etal.2019} find that the SMBH mass correlates with the X-ray quantities for their whole mixed-type sample of galaxies, and the scatter remains approximately constant across the investigated dynamical range. They interpret the correlations between the SMBH mass and the hot atmospheres as a consequence of the chaotic cold accretion (CCA) of the gas condensed out of the turbulent X-ray atmosphere onto the central SMBH, whereby sustaining a self-regulated feedback loop. The result is consistent with a numerical study by \cite{bassini.etal.2019} who used simulated samples of galaxy clusters obtained from zoom-in simulations to show that SMBH mergers are too rare to establish the observed correlations by means of the central-limit theorem. This line of interpretation embraces the scenario in which the correlations between the central SMBHs and their host galaxy's properties are a direct result of a delicate feedback-regulated interaction between the SMBH and its host galaxy (\citealt{silk.rees.1998,king.2003,churazov.etal.2005}).   
%In fact, the explanation falls into two main categories: whether it is of casual or non-casual nature. (To provide all relevant discussion also from simulations).

Theoretically, the properties of hot atmospheres are expected to be shaped by a complex interplay of processes, such as gravitational heating, radiative cooling, stellar and AGN feedback (see e.g. \citealt{werner.etal.2019} for a review), making the explanation for the origin of the observed correlations non-trivial. At the cluster mass and spatial scales, the thermodynamical properties of the ICM are to zeroth-order mainly determined by the total mass of the cluster according to the self-similar model (\citealt{kaiser.1986}), albeit the X-ray observations of the ICM reveal small deviations from the self-similar predictions (see \citealt{giodini.etal.2013} for a review). Moving toward the smaller scales of galaxies (i.e. integrating or mediating the X-ray signals across smaller galactocentric apertures and around lower-mass galaxies than BCGs), the properties of the observed atmospheric gas appear more susceptible to feedback processes (e.g. \citealt{kim.fabbiano.2015,goulding.etal.2016,lakhchaura.etal.2018,babyk.etal.2018}).

As a step toward better understanding the origin of the relationship between the mass of SMBHs and the X-ray properties of hot atmospheres, in this paper we systematically study such correlations from a theoretical and numerical perspective, from galactic to cluster scales. In particular, we inspect the correlations among X-ray gas properties (temperature and luminosity, in the $\sim 0.3-7$ keV range), SMBH masses, and total halo masses in different mass and spatial scale regimes and by varying the impact of astrophysical processes, such as feedback activity relative to gravitational physics. For this study, we employ data taken from the TNG100 run of the IllustrisTNG cosmological simulations (\citealt{marinacci.etal.2018,naiman.etal.2018,nelson.etal.2018,pillepich.etal.2018,springel.etal.2018}). These simulations self-consistently produce the large scale structure from the sub-galactic scales of galaxies to those of groups and clusters of galaxies. Furthermore, we use additional simulations with varying underlying physical models to quantify the effects of different astrophysical processes. In order to make a truthful comparison with X-ray data of galaxies, we carry out mock X-ray observations for the simulated galaxies. However, given their complexity and unknown nature, we cannot accurately reproduce the selection functions of the observed galaxy samples against which the IllustrisTNG outcome is placed. 

The goals of this paper are twofold: i) to inspect the relationships that emerge from a state-of-the-art galaxy formation model between the masses of SMBHs and the X-ray properties of the hot gaseous atmospheres around them across a wide range of mass and spatial scales, i.e. from galaxies to clusters of galaxies, and compare these to observations; and ii) to examine the effect of stellar and SMBH feedback on such scaling relations. 

The paper is organised as follows. In Section~\ref{sec:method}, we describe our methodology, including: description of the simulated data, procedures of mock X-ray analysis, and the X-ray observed samples. In Section~\ref{sec:3}, we present the predictions from the TNG100 simulation, including the growth of SMBHs, the correlations between SMBH masses with the X-ray properties of the hot atmospheres, and their radial dependence. There we also derive analytical scaling relations -- and compare them to the simulation results -- between the masses of SMBHs and the physical properties of the gas under the self-similar ansatz and the assumption that SMBH growth traces total halo mass growth. Section~\ref{sec:comparison_xray} presents a comparison with observational results. Then in Section~\ref{sec:theo} we explore the effects of stellar and SMBH feedback on the SMBH-hot atmospheres correlations. Finally we summarise the main results and conclude in Section~\ref{sec:conclusion}.

\section{Methodology}
\label{sec:method}
%%%%%%%%%%%%%%%%%%%%%%%%%%%%%%%%%%%%%%%%%%%%%%%%%%%%%%%%%%%%%%
\subsection{The TNG100 sample and TNG model variations}
\label{sec:tng}
%%%%%%%%%%%%%%%%%%%%%%%%%%%%%%%%%
%\ntnote{To smooth the text later!}
Here, we briefly describe the simulations as well as the key features of their underlying physical model. The sample utilised in this study is taken from the cosmological magnetohydrodynamical simulations IllustrisTNG\footnote{http://www.tng- project.org} (hereafter TNG, \citealt{marinacci.etal.2018,naiman.etal.2018,nelson.etal.2018,pillepich.etal.2018,springel.etal.2018}). The simulations are run with the {\sc arepo} code (\citealt{springel.2010}) and, to follow the evolution of the background Universe, a $\Lambda$-CDM cosmological model is used with cosmological parameters obtained based on the measurements from the Planck satellite (\citealt{planck.2016}): matter density $\Omega_{\rm m}=0.3089$, baryon density $\Omega_{\rm b}=0.0486$, dark energy density $\Omega_\Lambda=0.6911$, Hubble constant $H_0=67.74\ {\rm km}\ {\rm s}^{-1}\ {\rm Mpc}^{-1}$, amplitude of the matter power spectrum parameterized by $\sigma_8=0.8159$, and primordial spectral index $n_s=0.9667$.    

Galaxy formation in the TNG simulations (\citealt{weinberger.etal.2017,pillepich.etal.2018a}) is realised via a wide range of important astrophysical processes: primodial and metal-line radiative cooling; modelling of star formation and evolution, supernova feedback and chemical enrichment; and modelling of SMBH growth and feedback. 

The TNG simulations were performed in different flagship runs with varying resolution and simulated volumes. For this study, we employ the simulated data at $z=0$ from TNG100 because it is an optimal combination between having a reasonably good resolution ($m_{\rm baryon}=1.4\times10^{6}M_\odot$) and a sufficiently large simulated box with a cubic length of $(110.7\ \rm{Mpc})^3$. The size of the simulated box is comparable to the volume probed by current X-ray observations of massive galaxies in the local Universe (within $\sim100$ Mpc), while the relatively good resolution of TNG100 allows us to investigate the hot atmospheres from galactic ($\sim$ few kpc) to cluster scales ($\sim$ Mpc). The reader is referred to \citealt{truong.etal.2020} for a comparison of the X-ray properties of the gaseous haloes between TNG100 and the higher-resolution run TNG50 \citep{nelson.etal.2019a, pillepich.etal.2019}.

In addition to TNG100 galaxies, we employ a few model variation runs in which different physical recipes are implemented in comparison to the fiducial TNG model. These are performed at the same numerical resolutions as TNG100 on a cosmological volume of about 37 comoving Mpc a side and are labeled L25n512 \citep{pillepich.etal.2018a}. In particular, we will use a set of model variation runs where the stellar and SMBH feedback are altered compared to the fiducial model, as described in Section~\ref{sec:theo} and summarized in Table~\ref{tb2}.

Throughout the paper, we select simulated galaxies at $z=0$ with $M_{\rm BH}>0$ for the study of SMBH growth. We limit our X-ray analysis to a mass-selected sample of galaxies with $M_*>3\times10^{9}M_\odot$. We include both central and satellite galaxies, unless otherwise explicitly noted.

%%%%%%%%%%%%%%%%%%%%%%%%%%%%%%%%%%%%%%%%%%%%%
\subsubsection{Modelling of SMBH growth and feedback in TNG simulations}
%%%%%%%%%%%%%%%%%%%%%%%%%%%%%%%%%%%%%%%%%%%%%
Here, we briefly recall the key aspects of the SMBH model in the TNG simulations (see \citealt{weinberger.etal.2017,weinberger.etal.2018} for a full description). A SMBH seed mass of $1.18\times10^6M_\odot$ is assigned to a Friend-of-Friend ({\sc fof}) identified halo with a mass above $7.38\times10^{10}M_\odot$ that does not yet harbor a SMBH. The SMBH can then grow by gas accretion according to the Eddington-limited Bondi model (see  Eqs. 1-3 in \citealt{weinberger.etal.2018}) and by mergers with other SMBHs.

In the  TNG simulations, AGN feedback is modelled by a two-mode model, in which the SMBH can release feedback energy to the surrounding environment in the form of thermal (thermal mode) and kinetic (kinetic mode) energy. The amount of injected energy is determined by the accretion rate onto the SMBH (see Eqs. 7-9 and the associated efficiency parameters in \citealt{weinberger.etal.2017}). The difference between the two feedback modes is set by the accretion rate onto the SMBH for which the thermal mode is active at high accretion rates, while at low accretion rates the kinetic mode is switched on. More precisely, the kinetic mode occurs when the Eddington ratio ($\chi$) falls below the threshold:
\begin{equation}
\chi = \rm{min}\bigg[0.002\bigg(\frac{M_{\rm BH}}{M_{\rm pivot}}, 0.1\bigg)\bigg], \label{00}
\end{equation}
where $M_{\rm BH}$ is the SMBH mass and $M_{\rm pivot}=10^8 M_\odot$ is a characteristic SMBH pivot mass. The numerical values of Eq.~\ref{00} had been determined via a calibration process aimed at reproducing, among other galaxy statistics, realistic stellar properties such as the stellar mass function and the spatial extent of the stellar population at $z=0$ (see \citealt{weinberger.etal.2017,pillepich.etal.2018a} for details). 

\begin{figure*}
    \centering
    \includegraphics[width=0.49\textwidth]{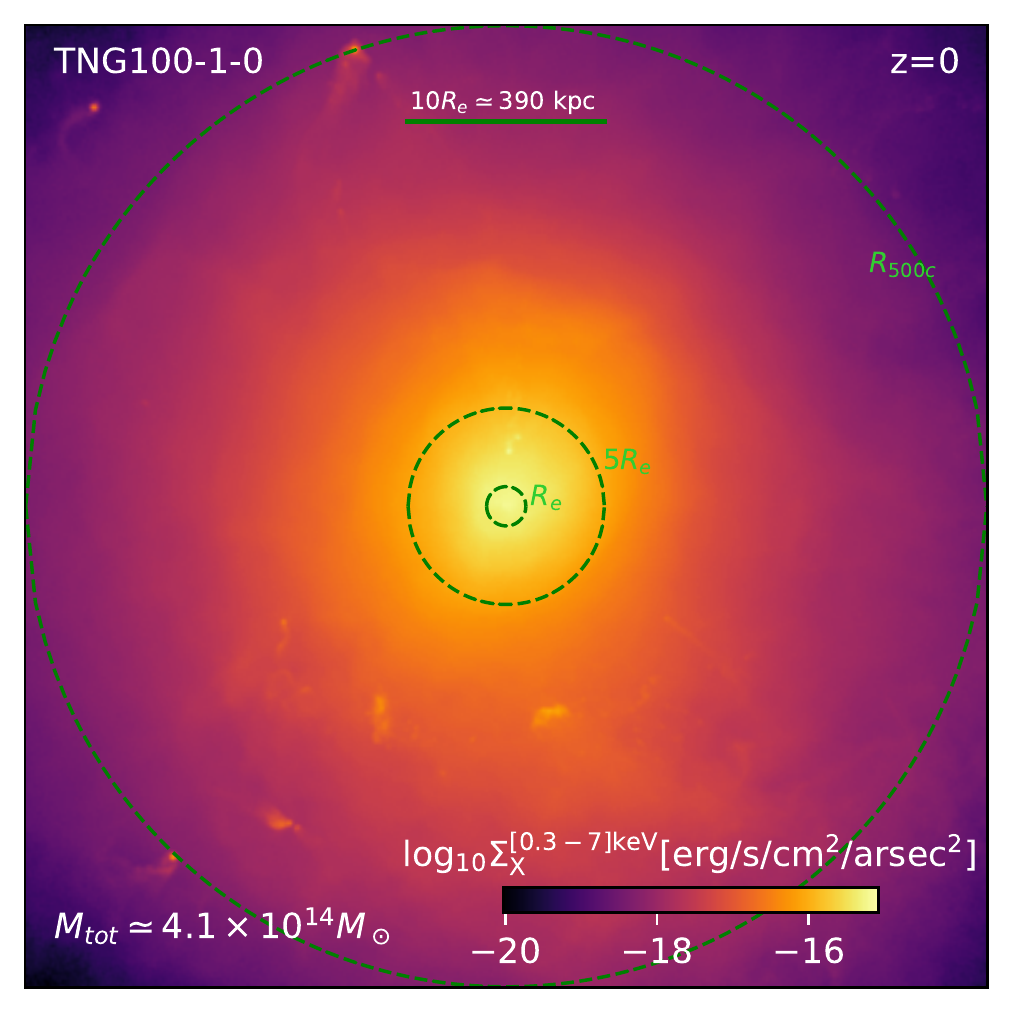}
    \includegraphics[width=0.49\textwidth]{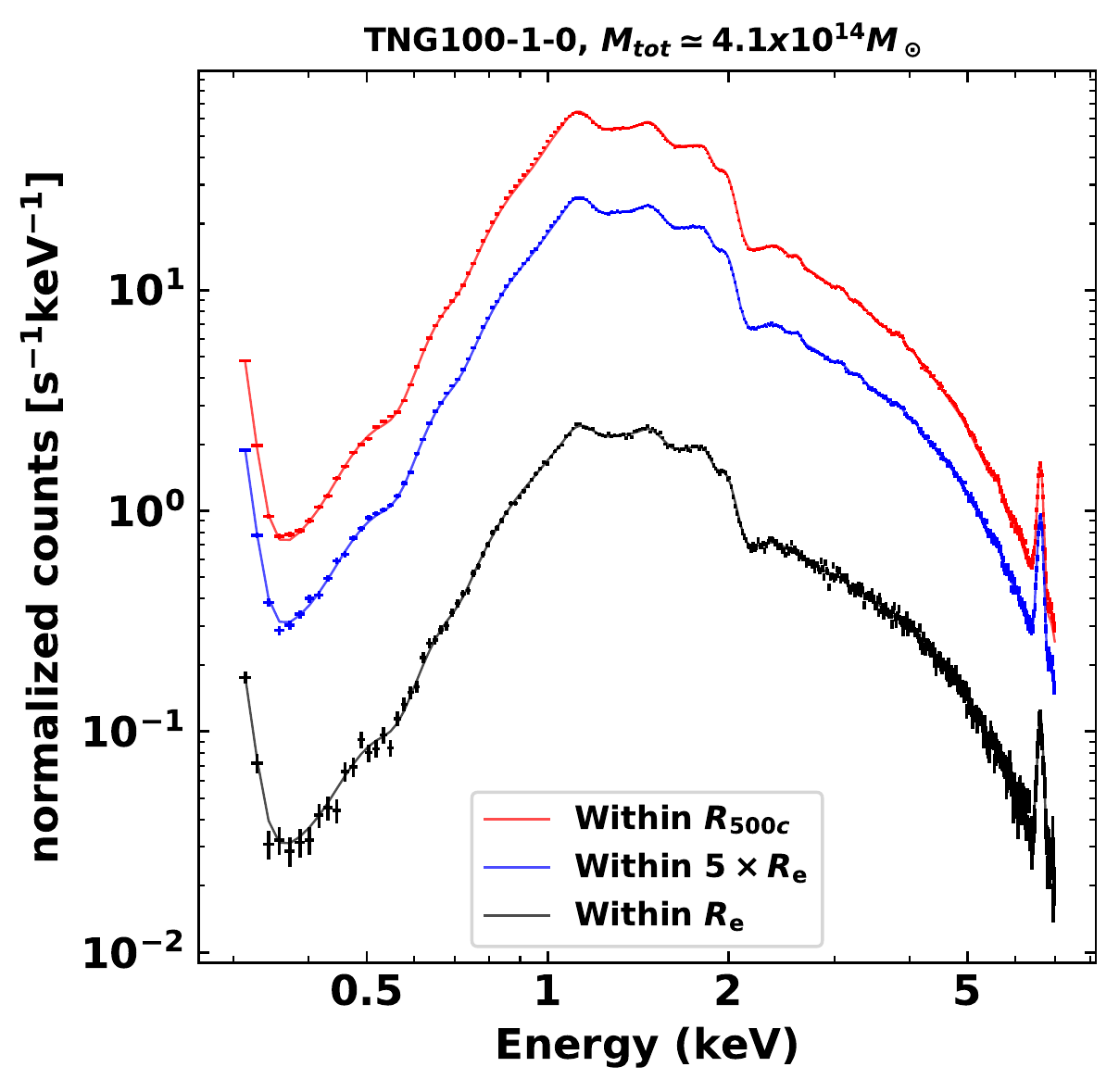}
    \caption{Illustration of the mock X-ray analysis for a TNG100 massive galaxy at $z=0$. {\it Left:} the X-ray surface brightness map of the galaxy within the cube $2R_{\rm 500c}\times2R_{\rm 500c}\times2R_{\rm 500c}$ ($R_{\rm 500c}\simeq1$ Mpc). The three circles mark the regions within $R_{\rm e}$, $5R_{\rm e}$, and $R_{\rm 500c}$, approximately indicating what in our jargon we refer to as the gas within the galactic bodies, within the gaseous coronae, and throughout the intrahalo/intragroup/cluster media. {\it Right:} The mock X-ray spectra extracted from the three regions as well as the corresponding best-fit curves obtaining from the fitting the spectra to a 1T APEC model (see text).} 
    \label{fig:0}
\end{figure*}

%%%%%%%%%%%%%%%%%%%%%%%%%%%%%%%%%%%%%%%%%%%%
\subsection{Computation of the properties of simulated galaxies}
%%%%%%%%%%%%%%%%%%%%%%%%%%%%%%%%%%%%%%%%%%%%
%%%%%%%%%%%%%%%%%%%%%%%%%%%%%%%%%%%%%%%%%%%%%%%%%
\subsubsection{Mock X-ray analysis}
\label{sec:mockxray}
%%%%%%%%%%%%%%%%%%%%%%%%%%%%%%%%%%%%%%%%%%%%%%%%%
To make a fair comparison with observations, we carry out a mock X-ray analysis of the simulated galaxies that resembles the procedure of analysing real X-ray data. The mock procedure consists of two steps: i) generating mock X-ray spectra on a galaxy-by-galaxy basis, obtained by summing up the spectra from individual gas cells; and ii) fitting the mock spectra to obtain for each galaxy the X-ray quantities of interest, such as gas temperature and luminosity (see \citealt{truong.etal.2020} for a detailed description of the method, here unchanged, and a discussion on relevant issues). 

For each galaxy, we start with gas cells that are selected by the {\sc subfind} algorithm, namely they are gravitationally bound to the host galaxy. We select non star-forming gas cells that lie within a cylindrical region with a radius of interest. The cylinder is centred at the most gravitationally bounded element of the galaxy, determined by {\sc subfind}, and it is randomly oriented, in our case along the z-axis of the simulation box, regardless of the galaxy configuration. The height of the cylinder is set to equal $10R_{\rm e}$, where $R_{\rm e}$ is the half-light radius (see the definition below), to account for projection effects. 
%\ntnote{This is motivated by the fact that in practice X-ray observers take all photons out to radii of $\sim3-5\ R_{\rm e}$ (e.g. \citealt{goulding.etal.2016,lakhchaura.etal.2019}), where the X-ray gaseous emission {\ap is expected to be} equal to the background emission {\ap AP: I do not understand: are we not talking about the choice of how many gas cells to integrate along the line of sight?}}.
To study the galactic X-ray emission, we make measurements for regions within radii of $R_{\rm e}$ and $5R_{\rm e}$ for both satellite and central galaxies. For central galaxies, we further measure the X-ray emission within $R_{\rm 500c}$, the spherical radius within which the mean density is 500 times greater than the critical density of the Universe (see detailed definition below in Section~\ref{sec:properties}). 

We note that our choice of selecting {\sc subfind}-based gas cells from simulations is aimed to investigate X-ray properties of gaseous atmospheres that are gravitationally bound to their host galaxy. This selection, for instance in the case of central galaxies, automatically excludes the X-ray emission coming from satellite galaxies that belong to the same halo, while for the case of satellite galaxies, it excludes the contribution from the intra-halo gas, which is typically gravitationally bound to the halo's central and most luminous galaxy. Excising the contribution from satellites for centrals and from the background for satellites is what is typically done in X-ray observations of galaxies in the local Universe, against which we compare in this paper. However, this does not correspond to what would be done in observational regimes with lower available spatial resolution. In addition, it is worth noting that the {\sc subfind}-based selection also excludes unbound, fast-moving gas elements driven by stellar or SMBH feedback. Those excluded gas elements could contribute non-negligibly to the total X-ray emission of the hot atmospheres, in particular in star-forming galaxies. We comment on how this selection affects our results at the end of Section~\ref{sec:comparison_xray} and provide a quantitative assessment in Appendix~\ref{sec:app1}.  

To facilitate the comparison with real X-ray observations of galaxies, a large fraction of which were obtained by the {\it Chandra X-ray Observatory} (see Section~\ref{sec:xraydata} below), the mock X-ray spectra are generated in a {\it Chandra}-like way. For each gas cell, mock spectrum is produced by using the {\sc fakeit} procedure implemented in XSPEC (\citealt{smith.etal.2001}), based on its thermodynamical properties: gas density, temperature, and metallicity, assuming a model for X-ray emission. In our study, we assume an absorbed single-temperature APEC model [‘wabs(apec)’]. The mock spectrum is generated for an exposure time of 100 ks without applying any background, and we assume Poisson statistical errors, based on the photon counts. A column density of $n_{\rm H}=10^{20}\ {\rm cm}^{-2}$ is applied for galactic absorption. The {\it Chandra}-like fashion is achieved by convolving the mock spectrum by the {\it Chandra} response files\footnote{We use response files for the default pointing of the 20th-cycle {\it Chandra} ACIS-S detector.} with a spectral resolution of 150 eV. The final mock spectrum of a galaxy within a certain radius is obtained by summing up the spectra created for all the gas cells that fall within the region of interest. 
It is then fitted to a single-temperature APEC model [‘wabs(apec)’] (1T) considering only the photon counts in the energy range of 0.3--7.0 keV (see \citealt{truong.etal.2020} for a discussion on single vs. two temperature model results).
%It is then fitted to either a single-temperature APEC model [‘wabs(apec)’] (1T) or a two-temperature model [‘wabs(apec + apec)’] (2T) considering only the photon counts in the energy range of 0.3--7.0 keV. 
For spectral fitting, we fix the gas metallicity to its pre-computed emission-weighted value\footnote{It is defined in a similar fashion as the emission-weighed temperature given in Eq.~\ref{eqn:2}, i.e. $Z_{\rm ew}=\frac{\sum_i \epsilon_i\times Z_i}{\sum_i\epsilon_i}$, where $Z_i$ is metallicity of the $i-th$ gas element and $\epsilon_i$ is its X-ray emission in the energy range 0.3--5.0 keV.}. For the Solar abundances we assume the values provided by \cite{anders.grevesse.1989}, but we have checked that using instead the element-by-element abundances that are directly output by the simulation affects the resulting X-ray luminosities and fitted temperatures by less than 5 per cent. The spectral fitting procedure provides best-fitting values for model parameters, i.e. the normalisation, which is proportional to the gas density squared, the gas temperature, as well as their associated uncertainties. The X-ray luminosity is derived based on the best-fit model in the energy range of 0.5--7.0 keV when comparing to observations and in the 0.3--5.0 keV band otherwise. 

To illustrate the procedure of our mock X-ray analysis, in Fig.~\ref{fig:0} we show an example of a massive central galaxy in TNG100. As shown in the left panel, we can study the system's hot atmospheres from galactic ($\sim R_{\rm e}$) to cluster scales ($\sim R_{\rm 500c}$). The X-ray emission is extracted from three different regions: within projected $R_{\rm e}$, $5R_{\rm e}$, and $R_{\rm 500c}$. The mock X-ray spectra associated with the regions are presented in the right panel, shown along with the corresponding best-fit 1T model of $\sim4.3$, $\sim3.6$, and $\sim3.1$  keV, from smaller to larger apertures for the integration, respectively.
  
\subsubsection{The definition of the ``X-ray detected'' sample} 
As thoroughly discussed in \citealt[][ see their Appendix A]{truong.etal.2020}, not all  mass-selected galaxies have reliable results obtained from the mock X-ray analysis described above. This is due to a combination of various factors: the sensitivity of the simulated instrument ({\it Chandra} ACIS-S), the choice for the exposure time (100 ks), the simulation resolution, and the physical condition of simulated systems, i.e. they may contain a too small amount of hot gas to be detected. As a result, a large fraction of the mass-selected sample produces too few photons in the considered energy range of 0.3--7 keV. To assess the goodness of the fitting results, we compare the obtained temperatures with their intrinsic value, such as the gas mass-weighted ($T_{\rm mw}$) and emission-weighted temperatures ($T_{\rm ew}$), given by:
\begin{equation}
    T_{\rm mw}=\frac{\sum_i m_{\rm g,i}\times T_i}{\sum_i m_{\rm g,i}}, \label{eqn:20}
\end{equation}
\begin{equation}
    T_{\rm ew}=\frac{\sum_i\epsilon_i\times T_i}{\sum_i\epsilon_i},\label{eqn:2}
\end{equation}
where $T_i$ ($m_{\rm g,i}$) is the temperature (mass) of the $i-th$ gas cell, and $\epsilon_i$ is its X-ray emission in the energy range 0.3--5 keV. We exclude systems with a best-fitting temperature ($T_{\rm X}$) that lies beyond the $3\sigma$ region around the average $T_{\rm X}-T_{\rm ew}$ relation. This selection is approximately corresponding to a lower limit of X-ray luminosity $L_{\rm X}(<R_{\rm e})\gtrsim5\times10^{37}\ {\rm erg/s}$. We are thus left with a subsample with reliable mock X-ray measurements, that is about $\sim34\%$ of the original mass-selected sample ($M_{*}>3\times10^{9}M_\odot$). We label this subsample as `X-ray detected' throughout the paper, comprising approximately 3500 galaxies in TNG100 at $z=0$.

%%%%%%%%%%%%%%%%%%%%%%%%%%%%%%%%%%%%%%%%%%%%%%%%%%%%%%
\subsubsection{Other properties of simulated galaxies}
\label{sec:properties}
%%%%%%%%%%%%%%%%%%%%%%%%%%%%%%%%%%%%%%%%%%%%%%%%%%%%%%
\begin{itemize}
    \item {\it Half-mass ($r_{1/2}$), half-light ($R_{\rm e}$), and $R_{\rm 500c}$ radii.} The stellar half-mass radius is defined as a spherical radius within which half of the galaxy's stellar mass is contained. Within the half-light radius half of the galaxy's stellar light is contained. We use the former to define mass measurements, e.g. the stellar mass, while the latter is more appropriate to compare with X-ray observations. For this work, we use the 2D circularised projected half-light radii estimated in K-band (\citealt{genel.etal.2017}; these measurements do not account for dust effects). Finally, to characterise the large cluster-scale properties, albeit only for central galaxies, we use $R_{\rm 500c}$, which is defined as the spherical radius within which the average matter density is 500 times the critical density of the Universe at a given redshift, i.e. it satisfies the relation: $M_{\rm 500c}=\frac{4}{3}\pi\times500\rho_{\rm crit,z=0}\times R_{\rm 500c}$.     
    \item {\it Galaxy stellar mass ($M_*$) and total halo mass ($M_{\rm tot}$).} The stellar mass of a galaxy is the stellar mass that is gravitationally bound and within twice the half-mass radius in 3D, i.e. $M_*(<2\times r_{1/2})$. The total dynamical mass of a galaxy, i.e. its total halo mass, is defined based on all particles that are gravitationally bound to that galaxy, according to the {\sc subfind} algorithm. Although the latter differs from the typical spherical-overdensity halo mass definitions such as $M_{\rm 500c}$ or $M_{200c}$, it is a very good proxy, with $M_{\rm tot}\propto M_{200c}$ with slope remarkably close to 1 throughout the mass range therein studied.
    \item {\it Flags of star formation status.} To characterise the star formation status of galaxies, we subdivide them, based on the relative difference in instantaneous star formation rate with respect to the star-forming main sequence at the corresponding stellar mass, into the following three categories (\citealt{pillepich.etal.2019}):
    \begin{itemize}
        \item Star-forming: $\Delta\log_{10}{\rm SFR}>-0.5$.
        \item Quenched: $\Delta\log_{10}{\rm SFR}\leq-1.0$.
        \item Green valley: $-1.0<\Delta\log_{10}{\rm SFR}\leq-0.5$.
    \end{itemize}
    Note that at $z=0$ (and only there) using either the instantaneous SFR values or SFRs integrated over e.g. the last 200 or 1000 million years affects negligibly the resulting quenched fractions (see e.g. Fig. 2 of \citealt{donnari.etal.2020b}).
    
    \item {\it Gas entropy.} For each $i-th$ gas element, its entropy ($K_i$) is defined as
    \begin{equation}
        K_i=\frac{k_{\rm B}T_i}{n_i^{2/3}}, \label{eqn:3}
    \end{equation}
where $k_{\rm B}$ is the Boltzmann constant, and $n_i$ is the number density of the gas cell. To estimate a representative value for a sample of gas elements, we use the emission-weighted estimation ($K_{\rm ew}$), similar to $T_{\rm ew}$ as given in Eq.~\ref{eqn:2}, for the entropy.
    \item {\it Virial temperature ($T_{\rm vir}$).} Under the assumption that the hot atmospheres are in virial equilibrium, their virial temprature can be approximated as:
\begin{equation}
\label{eq_tvir}
    T_{\rm vir}\equiv T_{\rm 200c}\simeq\frac{\mu m_{\rm p}GM_{\rm 200c}}{2k_{\rm B}R_{\rm 200c}}\simeq\frac{\mu m_{\rm p}GM_{\rm tot}}{2k_{\rm B}R_{\rm 200c}},
\end{equation}
where $\mu\simeq0.59$ is the mean molecular weight, $m_{\rm p}$ is the proton mass, $G$ is Newton's gravitational constant, $k_B$ is the Boltzmann constant. 
    \item {\it SMBH mass ($M_{\rm BH}$).} The SMBH mass of a galaxy is the total mass of all SMBHs that belongs to that galaxy at any given time, determined via {\sc subfind} algorithm. However, $M_{\rm BH}$ is completely dominated by the central massive SMBH, while the other SMBHs' mass fraction is negligible ($\sim10^{-5}$).   
\end{itemize}
%%%%%%%%%%%%%%%%%%%%%%%%%%%%%%%%%%%%%%%%%%
\subsection{The Observational Datasets}
\label{sec:xraydata}
%%%%%%%%%%%%%%%%%%%%%%%%%%%%%%%%%%%%%%%%%%
To compare with simulations, we use data from various X-ray observations of galaxies and clusters of galaxies in the local Universe ($D\lesssim150$ Mpc). It is worth noting that the selection of the existing samples of galaxies is not well-defined as it depends on the joint availability of SMBH mass measurements and X-ray data. We comment later in Section~\ref{sec:comparison_xray} on how this might affect the selection of the corresponding simulated sample for comparison with the observed data.
\begin{itemize}
    \item \cite{lakhchaura.etal.2019}: present a sample of 47 early-type galaxies observed by the {\it Chandra X-ray observatory}. The X-ray measurements are taken within $R_{\rm e}$ and $5R_{\rm e}$. The sample of galaxies is selected by matching a list of galaxies with available direct SMBH mass measurements in the literature (\citealt{kormendy.ho.2013,saglia.etal.2016,van.den.bosch.2016}) and the {\it Chandra} archive. The X-ray emission for each galaxy is modelled by a single-temperature APEC model.   
    \item \cite{gaspari.etal.2019}: compiled a larger sample of 85 systems, consisting mostly of early-type (i.e. ellipticals and S0s) galaxies, that have both direct SMBH mass measurement (primarily from \citealt{van.den.bosch.2016}) and X-ray observations taken from various X-ray instruments: {\it Chandra}, {\it ROSAT}, and {\it XMM-Newton}. The X-ray measurements are available for various radii ranging from galactic (e.g. within $R_{\rm e}$) to cluster scale (e.g. within $R_{\rm 500c}$). The X-ray spectra are fitted with an APEC model.
    \item \cite{bogdan.etal.2018}: studied a sample of 17 brightest-group/cluster galaxies (BGGs/BCGs) that have X-ray observations from {\it XMM-Newton} and direct SMBH mass measurements taken from \cite{kormendy.ho.2013}. For each system, the X-ray spectrum is obtained for a projected region within $R_{\rm 500c}$ and is fitted using an APEC model. 
\end{itemize}
%%%%%%%%%%%%%%%%%%%%%%%%%%%%%%%%%%%%%%%%%
\subsection{Fitting Methods}
\label{sec:fitting_method}
%%%%%%%%%%%%%%%%%%%%%%%%%%%%%%%%%%%%%%%%%%
Throughout the paper, to characterise the SMBH scaling relations, the data is fitted to a single power law in log space via a linear fitting method from the IDL routine linmix err.pro \citep{kelly.2007}:

\begin{equation}
    \log_{10}\bigg(\frac{M_{\rm BH}}{10^8 M_\odot}\bigg)= {\rm Normalization}+{\rm Slope}\times\log_{10}\bigg(\frac{F_{\rm X}}{F_0}\bigg), \label{eqn:10}
\end{equation}
where $F_{\rm X}$ stands for any galaxy, gaseous atmosphere or halo property, and $F_0$ is the corresponding pivotal value. 
This method adopts a bayesian approach to linear fitting, in which it treats the normalisation, the slope, and the intrinsic scatter as free parameters and they are estimated via the Monte Carlo Markov Chain method.

By construction, since lower-mass systems are more frequent in volume-limited samples and since the galaxy sets from cosmological simulations used here are volume limited, the results of our fitting procedure is more strongly weighted by the data points at the low-mass end, sometimes resulting in best-fit curves that do not well represent the high-mass end data. This does not mean that a single power law could not be found to describe to a reasonable approximation all the data points across the entire inspected mass scale; however, it also suggests a marginal change of slope with mass of the scaling relations under scrutiny.

Throughout the paper, we will quantify the significance of the correlation between two variables via the Pearson ($r_{\rm P}$) and Spearman ($\rho_{\rm S}$) correlation coefficients. For the sake of reference, a correlation with absolute values lying in the range: $[0.8, 1.0]$, $[0.5, 0.8]$, $[0.3, 0.5]$, and $[0.0, 0.3]$, is commonly considered as  strong, moderate, weak, and absent, respectively.   

%%%%%%%%%%%%%%%%%%%%%%%%%%%%%%%%%%%%%%%%%%%%%%%%%%%%%%%%%%%%%%%%%%
\section{SMBH--Hot atmospheres correlations in TNG100: from galactic to cluster scales}
%\section{SMBH--hot atmospheres relationships in the TNG100 simulation}
\label{sec:3}
%%%%%%%%%%%%%%%%%%%%%%%%%%%%%%%%%%%%%%%%%%%%%%%%%%%%%%%%%%%%%%%%%%%%
%%%%%%%%%%%%%%%%%%%%%%%%%%%%%%%%%%%%%%%%%%%%%%%%%%%%%%%%%%%%%%%%%%%%%%%%%%%%%%%%%%
In this Section, we present a quantification of the relationships between the mass of SMBHs and the X-ray properties of the hot gaseous atmospheres as they emerge from the TNG100 simulation. When it comes to studying SMBH-galaxy co-evolution, it is instructive to first inspect how SMBHs grow in the TNG context, for it reveals the fundamental relationship between SMBH mass and galaxy or halo mass.
However, before doing so and in order to facilitate the physical understanding of the correlations predicted by the simulations or empirically derived, we first recall the solution to the problem at hand within the framework of the self-similar model. 

%, as anticipated in the self-similar model, {\ap at least in selected mass regimes}.
%, SMBH relations with other galaxy properties, e.g. gaseous X-ray properties, could be derived. %\ntnote{to provide some statistics for the studied sample, i.e. the total number of galaxies, the number of X-ray detected!} 

%%%%%%%%%%%%%%%%%%%%%%%%%%%%%%%%%%%%%%%%%%%%%%%%%%
\subsection{Self-similar predictions for the SMBH--hot atmosphere scaling relations}
\label{sec:self-similar}
%%%%%%%%%%%%%%%%%%%%%%%%%%%%%%%%%%%%%%%%%%%%%%%%%%%%%%%%%%%%%%%%%%%
 According to the self-similar model (\citealt{kaiser.1986}), the thermodynamics of the hot atmospheres are determined by the gravitational potential well of the underlying dark matter halo (see \citealt{giodini.etal.2013} for a thorough review). 

The X-ray temperature of the gas in a collapsed halo can be related to the halo mass assuming virial equilibrium (see also Eq.~\ref{eq_tvir}):
\begin{equation}
    T_{\rm X}\sim T_{\rm vir}\propto\frac{GM_{\rm vir}}{R_{\rm vir}}\propto M_{\rm tot}^{2/3}, \label{eqn:4}
\end{equation}
where $M_{\rm vir}$ and $R_{\rm vir}$ are the virial mass and radius, respectively, and where the last approximation can be taken because of the linear relationship between total gravitationally-bound halo mass and virial mass.

For the X-ray luminosity, assuming the X-ray emission is mainly driven by thermal bremsstrahlung, $L_{\rm X}$ is given by:
\begin{equation}
    L_{\rm X}\propto f_{\rm gas}^2T_{\rm}^{1/2}M_{\rm tot}\propto M_{\rm tot}^{4/3}, \label{eqn:5}
\end{equation}
where $f_{\rm gas}\equiv\frac{M_{\rm gas}}{M_{\rm tot}}$ is the gas mass fraction and it is constant in the self-similar scenario.

In the original model of \cite{kaiser.1986}, a relation between $M_{\rm BH}-M_{\rm tot}$ is not provided. However, we can derive this relation based on the assumption that the SMBH mass traces the halo mass:
\begin{equation}
    M_{\rm BH}\propto M_{\rm tot}. \label{eqn:6}
\end{equation}
In fact, a linear relationship between SMBH and halo masses has been widely suggested in the literature, at least at the highest-mass end. For example, \cite{jahnke.maccio.2011} showed that in a hierarchical cosmological scenario, both SMBHs and haloes grow simultaneously via mergers and a linear relation with slope equal to 1 is expected as a natural consequence of the central limit theorem in a scenario where SMBHs grow {\it only} via BH-BH merging.

From Eqs.~\ref{eqn:4},~\ref{eqn:5}, and \ref{eqn:6}, one can derive a ``self-similar'' (and ``merger-driven'') prediction for the SMBH mass -- gas atmosphere X-ray scaling relations as follows:
\begin{equation}
M_{\rm BH}\propto T_{\rm X}^{3/2},\\ \label{eqn:8}
\end{equation}
\begin{equation}
M_{\rm BH}\propto L_{\rm X}^{3/4}. \label{eqn:9}
\end{equation}
It is worth emphasising that the above scaling relations are derived without assuming any causal interaction between SMBHs and the gaseous atmospheres. In other words, within the self-similar framework, these scaling relations are the results of the gravitational collapse of (also gaseous) haloes within a hierarchical growth of structure and are simply the reflection of the common relationships of $M_{\rm BH}$, $T_{\rm X}$, and $L_{\rm X}$, respectively, with total halo mass.

%%%%%%%%%%%%%%%%%%%%%%%%%%%%%%%%%%%% Fig%%%%%%%%%%%%%%%%%%%%%%%%%%%%%
\begin{figure*}
    \centering
    \includegraphics[width=0.99\textwidth]{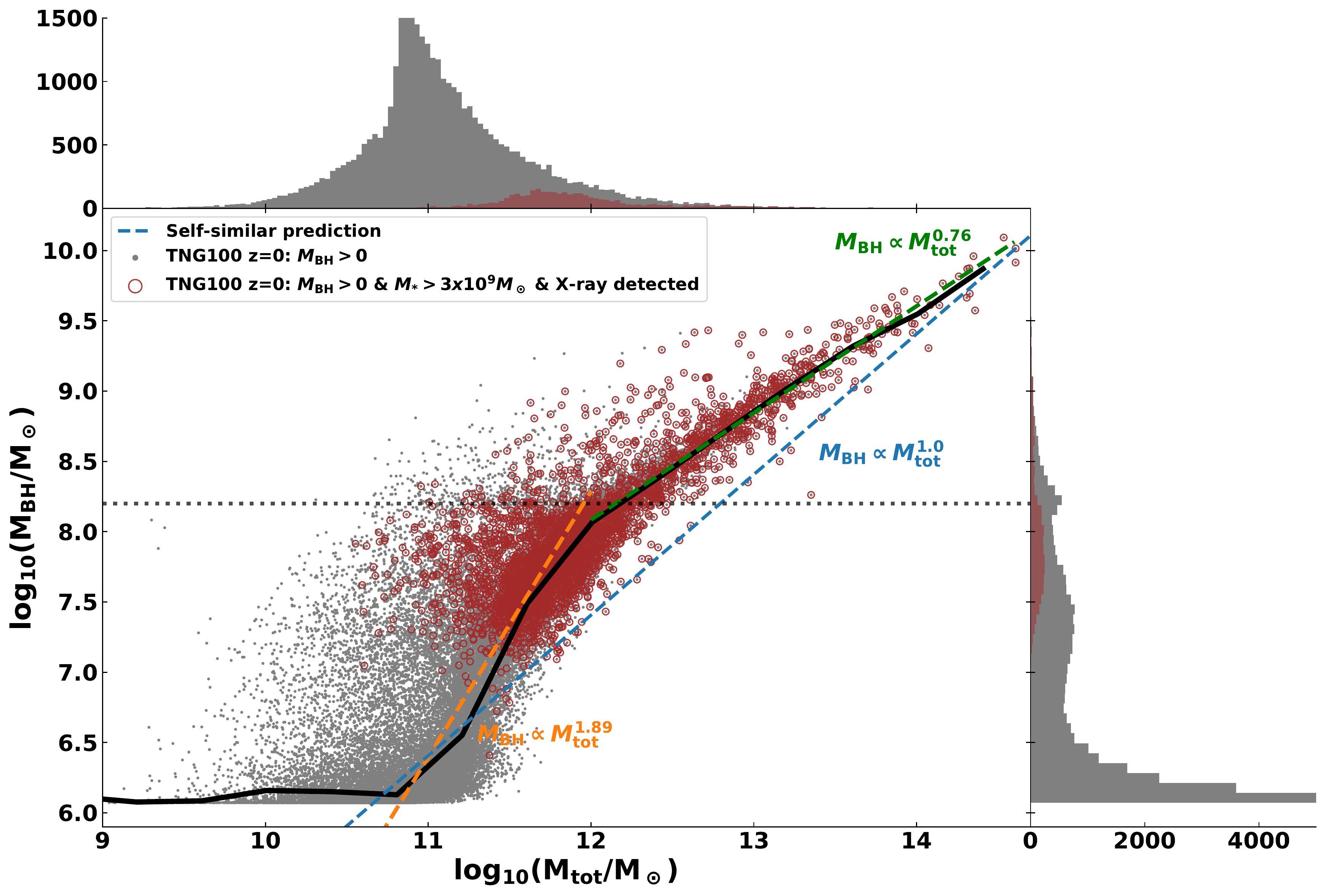}
    \caption{The growth of SMBHs in the TNG100 simulation and potential constraint from X-ray observations. {\it Main panel:} SMBH mass vs. total halo mass shown for the overall galaxy sample with $M_{\rm BH}>0$ (in grey) and the subsample with reliable X-ray measurements (in red) at $z=0$. The solid line represents the median relation for the overall sample. The horizontal dotted line approximately marks the SMBH mass scale ($M_{\rm BH}\sim10^{8.2}M_\odot$) where the third phase of SMBH growth starts, while the dashed lines serve as guidelines for the curve's steepness. By self-similar prediction (dashed blue line) here we indicate the SMBH--halo mass relation that could emerge through a SMBH growth exclusively driven by mergers.{\it Sub-panels:} histograms of $M_{\rm BH}$ and $M_{\rm tot}$.}  
    \label{fig:1}
\end{figure*}
%%%%%%%%%%%%%%%%%%%%%%%%%%%%%%%%%%%%%%%%%%%%%%%%%%%%%%%%%%%%%%%%%%%%%%%
\begin{figure*}
    \centering
    \includegraphics[width=0.9\textwidth]{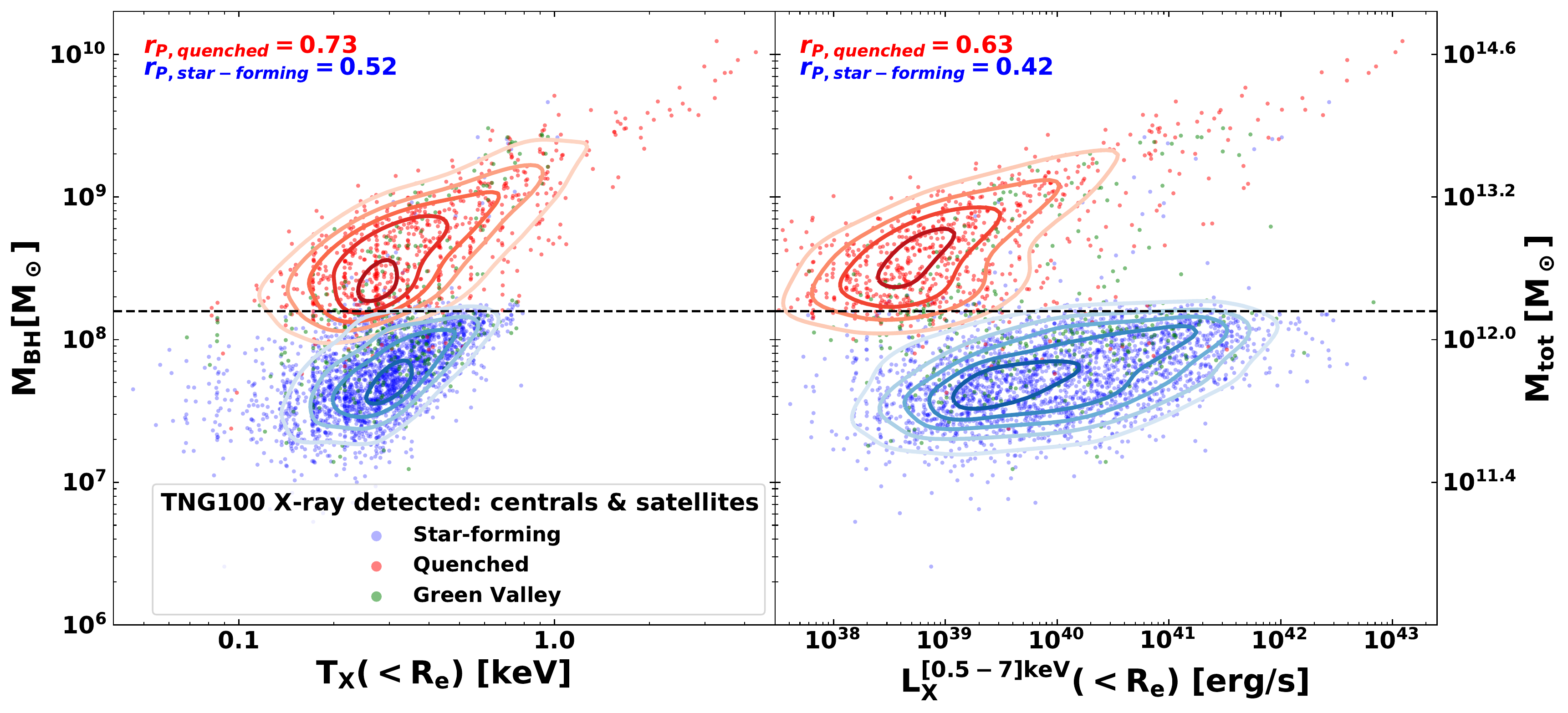}
    \includegraphics[width=0.9\textwidth]{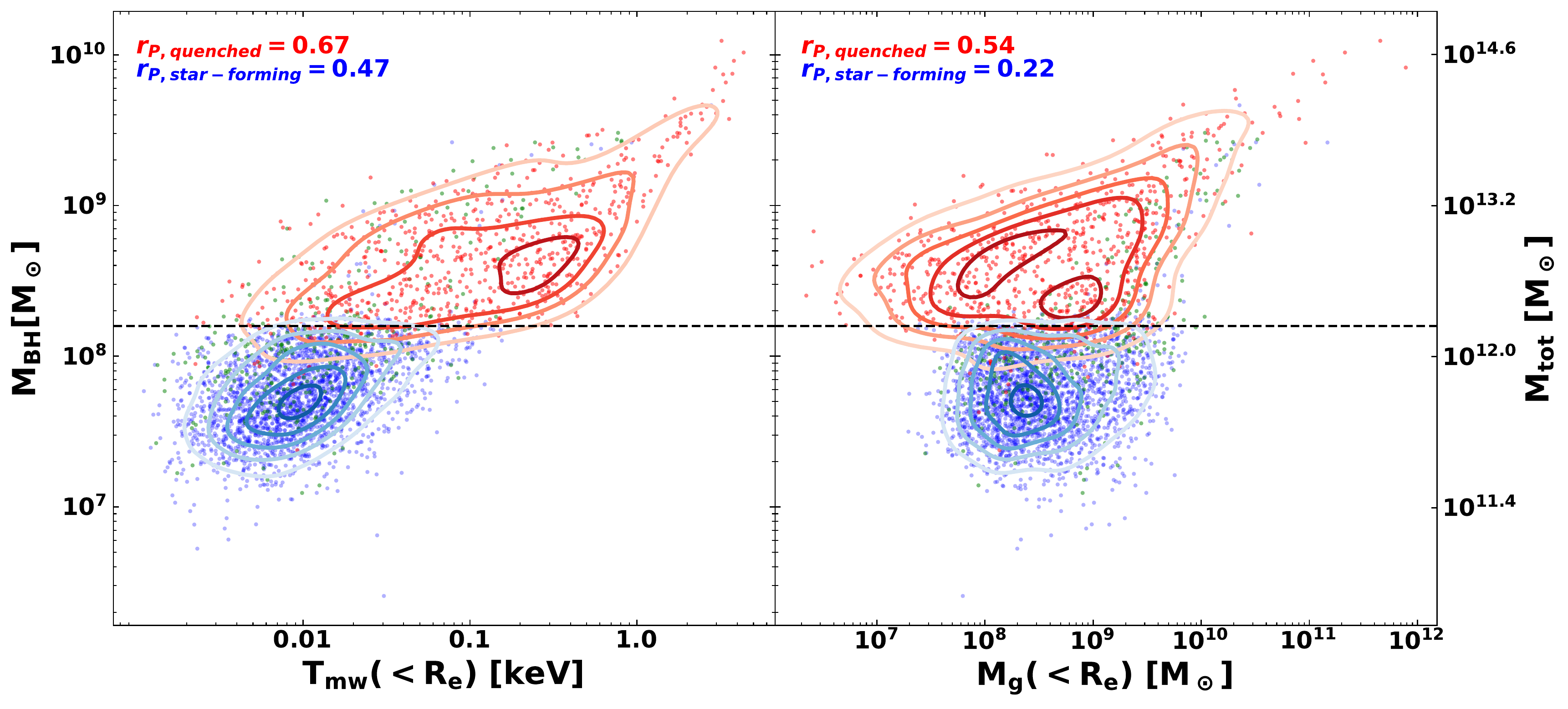}
    \includegraphics[width=0.9\textwidth]{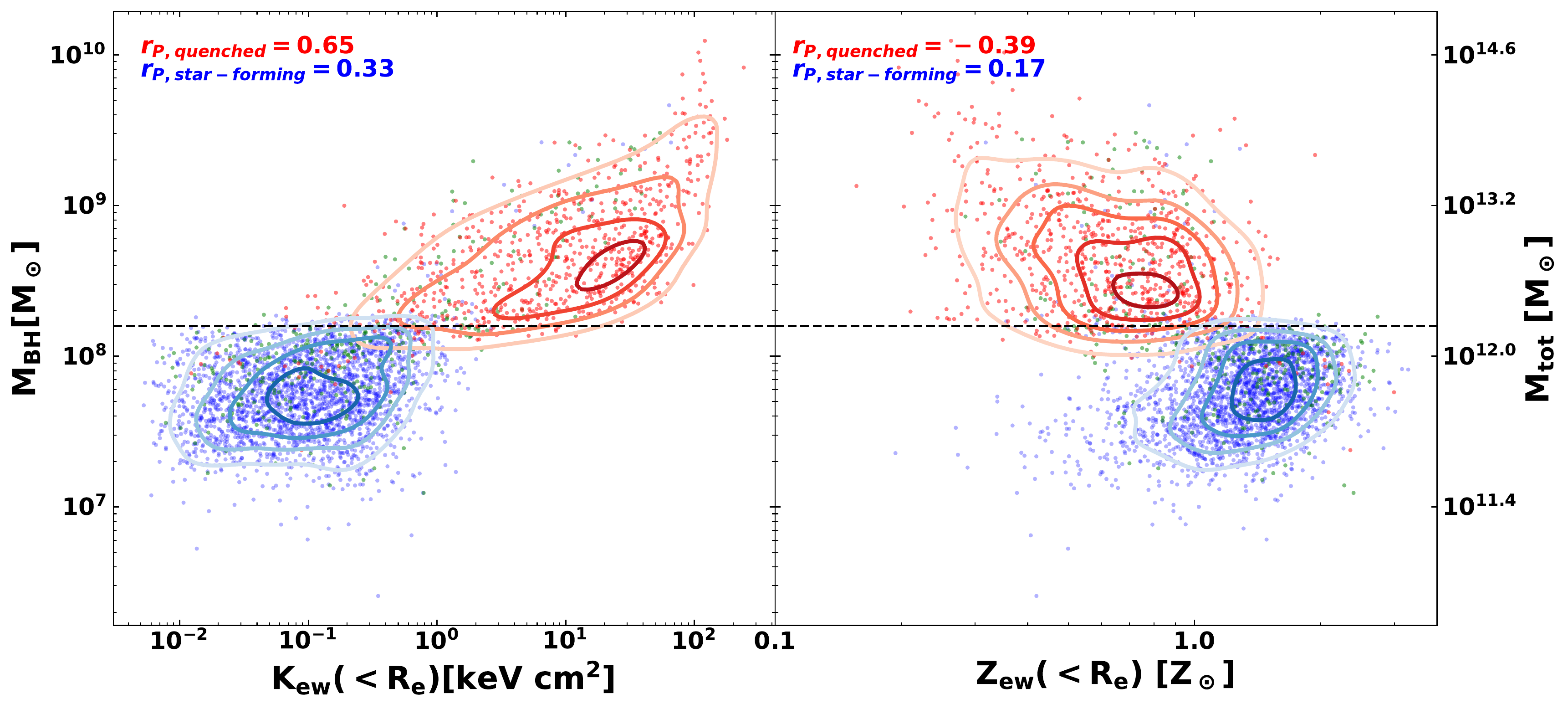}
    \caption{The correlations between SMBH mass and X-ray properties of the hot atmospheres, measured within $R_{\rm e}$, for different galaxy populations in TNG100 at $z=0$. It is shown the SMBH mass in relation with the gaseous X-ray temperature ($T_{\rm X}$), X-ray luminosity ($L_{\rm X}$), gas mass-weighted temperature ($T_{\rm mw}$), total gas mass ($M_{\rm g}$), emission-weighted entropy ($K_{\rm ew}$), and emission-weighted metallicity ($Z_{\rm ew}$). The red and blue contours specify the loci of quenched and star-forming populations, respectively. The horizontal dashed line marks $M_{\rm BH}\sim10^{8.2}M_{\odot}$ where approximately is the boundary line between quenched and star-forming populations.The values of the Pearson correlation coefficients for each populations are stated at the top-left corner of each plot. The rightmost y-axis represents the median value of $M_{\rm tot}$ according to the median relation $M_{\rm BH}-M_{\rm tot}$ as shown in Fig.~\ref{fig:1}.} 
    \label{fig:2}
\end{figure*}

%%%%%%%%%%%%%%%%%%%%%%%%%%%%%%%%%%%%%%%%%%%%%%%%%%%%%%%%%%%%%%%%%%%%
\subsection{$M_{\rm BH}-M_{\rm tot}$ relation at $z=0$}
\label{sec:bhgrowth}
%%%%%%%%%%%%%%%%%%%%%%%%%%%%%%%%%%%%%%%%%%%%%%%%%%%%%%%%%%%%%%%%%%%
 To have an approximate picture of the SMBH growth in the TNG simulations as a function of halo mass, we can examine the relationship between SMBH mass ($M_{\rm BH}$) and its total halo mass ($M_{\rm tot}$) at the present time. 
 
 Fig. \ref{fig:1} presents the $M_{\rm BH}-M_{\rm tot}$ relation for the TNG100 sample at $z=0$ consisting of 36941 galaxies with $M_{\rm BH}>0$. On top of that, we also show a subsample of 3513 galaxies for which X-ray measurements of their hot atmospheres can be reliably made with 100ks exposure on {\it Chandra}, according to the mock X-ray analysis described in Section~\ref{sec:mockxray}, in order to provide potential X-ray constraints on the $M_{\rm BH}-M_{\rm tot}$ relation.

Assuming that the average system evolves in time along the $z=0$ relation of Fig. \ref{fig:1}, the growth of SMBHs in TNG can be approximately divided into 3 distinct regimes:
\begin{itemize}
    \item {\it First} regime - $M_{\rm tot}\lesssim 10^{11}M_\odot$: SMBHs experience almost no growth at all and their mass remains close to the seed mass, i.e. $M_{\rm BH}\sim10^{6}M_\odot$.
    \item {\it Second} regime - $10^{11}\lesssim M_{\rm tot} \lesssim 10^{12}M_\odot$: SMBHs undergo a fast-growing period in which their masses increase by more than 3 orders of magnitudes (approximately $M_{\rm BH}\propto M_{\rm tot}^{1.89}$). 
    \item {\it Third} regime - $M_{\rm tot}\gtrsim10^{12}M_\odot$: the SMBH growth rate starts to decline and SMBH mass scales with total halo mass approximately as $M_{\rm BH}\propto M_{\rm tot}^{0.76}$.    
\end{itemize}
%%%%%%%%%%%%%%%%%%%%%%%%%%%%%%%%%%%%%%%%%%%%%%%%%%%%%%%%%%%%%%%%%%
In the {\it first} regime, SMBHs have just been seeded, thereby their mass is close to the seed mass. In addition, in this regime where they reside in low-mass (mostly) star-forming galaxies, SMBHs may experience difficulties in growing also due to stellar activity, e.g. supernova feedback, which can prevent the accumulation of cold gas in the central regions. This can be clearly seen when comparing with simulations without stellar feedback, as we discuss in greater detail later in this paper (Section~\ref{sec:theo}) and as already suggested or demonstrated via numerical models by a few authors \citep[e.g.][]{dubois.etal.2015,bower.etal.2017, pillepich.etal.2018a, habouzit.etal.2020}. 
% AP: see Habouzit, 2006.10094v1 and at her page 20 for citations
In more massive galaxies, the gravitational potential well is sufficiently deep to hold on to hot atmospheres, which inhibit outflows driven by supernovae feedback. Thus in the {\it second} regime, possibly thanks to the very formation of the hot atmospheres, SMBHs can grow fast via (cold) gas accretion. During this period, the SMBH growth is regulated by the depth of the galaxy's gravitational potential well, namely SMBHs cannot grow too fast otherwise their feedback would disrupt the accretion flows (see e.g. Fig. 6 in \citealt{terrazas.etal.2019}, also \citealt{booth.schaye.2010}). 

The fast-growing period is terminated in TNG at $M_{\rm tot}\sim10^{12}M_\odot$ where SMBHs reach the mass of $M_{\rm BH}\sim10^{8.2}M_\odot$. In TNG, this SMBH mass scale is associated, albeit indirectly and via the interplay with the evolution through cosmic epochs of structure formation and with other astrophysical mechanisms, to the pivot mass where SMBHs can switch from thermal to kinetic feedback mode (see Eq.~\ref{00}). Previous studies (\citealt{weinberger.etal.2018,davies.etal.2019,terrazas.etal.2019, zinger.etal.2020}) showed that in the TNG model the kinetic feedback plays a crucial role in quenching star formation by expelling gas out of the central regions of galaxies and by heating the gas up. SMBH feedback can thus prevent further gas accretion, thereby truncating the SMBH growth in massive galaxies, in a self-regulating fashion. In this high-mass regime within the TNG simulations, it has been explicitly demonstrated by \cite{weinberger.etal.2018} that the major channel of SMBH growth is via mergers with other SMBHs, which are following the mergers of their host haloes and galaxies. In particular, as it can be seen in their Fig. 7, SMBH mergers greatly dominate the growth of SMBHs for $M_{\rm BH(z=0)}\gtrsim10^{8.5} (10^{9.5})M_\odot$ at $z\lesssim1~ (2)$. It should be noted, however, that even in this regime, some level of gas accretion onto the SMBH does occur and the kinetic SMBH feedback remains the dominant non-gravitational heating process in TNG galaxies, as demonstrated by \cite{weinberger.etal.2018} in their Fig. 1. Notably, it is in this third regime of SMBH growth that we see in TNG the most significant correlation between the SMBH mass and the total halo mass ($r_{\rm P}\simeq0.87$) with a relatively small scatter ($\sigma_{M_{\rm BH}}\simeq0.14$ dex). The presence in the Universe of a $M_{\rm BH}-M_{\rm tot}$ scaling relation with a decreasing scatter toward the high-mass end could thus be a natural consequence of the hierarchical assembly of the large scale structure, in line with the expectations of the ``dry'' merger scenario  (\citealt{jahnke.maccio.2011} , see dashed blue line in Fig.~\ref{fig:1}).

It is important to note that a slope of the $M_{\rm BH}-M_{\rm tot}$ relation that differs from 1 above a certain mass scale does not mean that the growth of SMBHs is not dominated by mergers in that regime. As it can be seen in Fig.~\ref{fig:1}, in TNG the slope of the $M_{\rm BH}-M_{\rm tot}$ scaling relation in the third phase is shallower than the unit value predicted by \cite{jahnke.maccio.2011} within the merger scenario ($0.78$ versus $1$), even though we know that TNG SMBHs have grown in the greatest part via mergers at those mass scales. This deviation could be attributed to the fact that, prior to entering the third phase, SMBHs gain some extra mass via gas accretion with respect to the predicted linear $M_{\rm BH}-M_{\rm tot}$ relation by pushing upward the relation at the lower-mass end and by therefore tilting the resulted relation\footnote{The simulations employed in \cite{jahnke.maccio.2011} do not consider gas accretion at all and SMBHs in their model can grow {\it solely} via mergers with other SMBHs throughout cosmic epochs.}. 

Importantly, from  Fig.~\ref{fig:1} it can be seen that galaxies with $M_{\rm BH}\gtrsim10^{7}$ ($M_{\rm tot}\gtrsim10^{11}M_\odot$) start to host X-ray bright, detectable atmospheres (gray dots versus red circles in Fig.~\ref{fig:1}), thereby making X-ray observations a potential window to probe the $M_{\rm BH}-M_{\rm tot}$ relation in the third and in a part of the second phase of SMBH growth ($M_{\rm tot}\gtrsim10^{11.5}M_\odot$). We note that even though here we limit the X-ray measurements to TNG100 galaxies with $M_{*}>3\times10^{9}M_\odot$, we have verified using the higher resolution run TNG50 that there is no significant gain in trying to observe, and thus increase the number of, lower-mass galaxies with reliable X-ray measurements below this mass threshold (see also \citealt{truong.etal.2020}).   

%%%%%%%%%%%%%%%%%%%%%%%%%%%%%%%%%%%%%%%%%%%%%%%%%%%%%%%%%%%%%%%%%%%%
\subsection{SMBH -- gaseous halo connection}
\label{sec:bhxray}
%%%%%%%%%%%%%%%%%%%%%%%%%%%%%%%%%%%%%%%%%%%%%%%%%%%%%%%%%%%%%%%%%%%
%%%%%%%%%%%%%%%%%%%%%%%%%%%%%%%%%%%% Fig%%%%%%%%%%%%%%%%%%%%%%%%%%%%%
As shown in previous TNG studies, SMBHs feedback plays a central role in determining the thermodynamical properties of the hot atmospheres (\citealt{nelson.etal.2018b,truong.etal.2020, zinger.etal.2020}) as well as in quenching star formation in massive galaxies (\citealt{weinberger.etal.2017,nelson.etal.2018,davies.etal.2019,terrazas.etal.2019, donnari.etal.2020a}). In this Section we explore the relationship between SMBHs and the hot atmospheres by quantifying the correlation between the mass of SMBHs and a number of physical properties of the surrounding halo gas that determine its X-ray manifestations. 

Fig.~\ref{fig:2} shows SMBHs masses against X-ray temperature ($T_{\rm X}$), X-ray luminosity ($L_{\rm X}$), gas mass-weighted temperature ($T_{\rm mw}$), gas content represented by total gas mass ($M_{\rm g}$ -- excluding star-forming gas), emission-weighted estimations of gas entropy ($K_{\rm ew}$), and metallicity ($Z_{\rm ew}$). For this task, all the gas quantities are measured within $R_{\rm e}$, so within galactic spatial scales, where feedback are expected to leave their most prominent impact. Furthermore, the relations are examined with respect to the star formation status of the host galaxy subdivided in three categories as described in Section~\ref{sec:properties}: star-forming, green valley, and quenched galaxies in blue, green, and red data points, respectively. We also report the values of the Pearson correlation coefficient to quantify the significance of the SMBH correlations for the two major populations of quenched and star-forming galaxies. 

%%%%%%%%%%%%%%%%%%%%%%%%%%%%%%%%%%%%%%%%%%%%%%%%%%
\subsubsection{Effects of SMBH feedback on  the gas properties of quenched vs. star-forming galaxies}
%%%%%%%%%%%%%%%%%%%%%%%%%%%%%%%%%%%%%%%%%%%%%
It is clear from Fig.~\ref{fig:2} that the SMBH mass is closely linked to the separation between the two major populations of quenched and star-forming galaxies. The two populations split sharply at the SMBH mass of $M_{\rm BH}\sim10^{8.2}M_\odot$, above and below which galaxies exhibit very different gas thermodynamics. In brief, above this mass range quenched galaxies contain hotter, higher entropy, and metal-richer atmospheres, although with smaller gas mass than their star-forming counterparts (\citealt{davies.etal.2019,zinger.etal.2020}). This difference results in an observable diversity in the X-ray luminosity between star-forming and quenched galaxies, in which the latter are about an order of magnitude {\it less} X-ray luminous than the former, at fixed galaxy/SMBH/halo mass (\citealt{truong.etal.2020,oppenheimer.etal.2020}). 

The manifest and observationally-testable impact of SMBH feedback onto the gas properties of galaxies, and differently so for star-forming and quenched galaxies, is a key novel result of the TNG simulations, discussed in detail in previous papers and here summarized as follows. According to the TNG model, the mass scale $M_{\rm BH}\sim 10^{8.2}M_\odot$ signals the transitioning from a population of SMBHs mostly injecting thermal energy to a population of SMBHs mostly releasing feedback via kinetic, SMBH-driven winds (see e.g. Fig. 9 of \citealt{terrazas.etal.2019} and also Eq.~\ref{00}). The SMBH kinetic feedback has been demonstrated being the primary mean of star formation quenching in TNG central massive galaxies, via both ejective and preventative effects: the SMBH kinetic feedback efficiently expels gas out of the inner regions of galaxies while simultaneously heating up the surrounding gas, increasing its cooling times, and thereby preventing it from fueling future star formation (\citealt{davies.etal.2019, zinger.etal.2020}). The reduction of the gas content in quenched galaxies is the main cause for the diversity in X-ray luminosity between star-forming and quenched galaxies, as shown in \citealt{truong.etal.2020} (see also \citealt{oppenheimer.etal.2020}). Nonetheless, it should be noted that the separation could also be partly contributed by the SMBH thermal mode feedback (and possibly stellar feedback), in particular in the mass range $M_*\sim10^{10-10.5}M_\odot$, where SMBH thermal feedback is the dominant heating channel (\citealt{weinberger.etal.2018}) and hence possibly capable of boosting the X-ray emission of star-forming galaxies.

In TNG, above $M_{\rm BH}\sim10^{8.2}M_\odot$, the kinetic feedback mode is the dominant non-gravitational heating process (see Fig. 1 in \citealt{weinberger.etal.2018}): this is the case when averaging across galaxy populations -- i.e. despite the fact that sporadically individual galaxies may undergo episodes of SMBH thermal mode energy injection. Furthermore, the dominance of the SMBH kinetic feedback extends to progressively larger redshifts for higher $z=0$ SMBH/galaxy masses. In such regime, the SMBH kinetic feedback in TNG mainly helps compensating the radiative cooling of gas within and around galaxies, thus keeping low their star formation rates (\citealt{zinger.etal.2020}). In this high-mass regime, while the SMBH kinetic feedback plays a significant role in determining the thermodynamics of the hot atmospheres, its effect becomes progressively less important in massive galaxies/haloes in comparison to gravitational heating processes such as shock heating of infall gas or merger activities. This issue will be discussed thoroughly in Sections 3.4-3.6.

%%%%%%%%%%%%%%%%%%%%%%%%%%%%%%%%%%%%%%%%%%%%
\subsubsection{Correlations of SMBH mass with X-ray and physical gas properties: quenched vs. star-forming galaxies}
\label{sec:BH-Xray_corr}
%%%%%%%%%%%%%%%%%%%%%%%%%%%%%%%%%%%%%%%%%%%%
We find that for the population of quenched galaxies the SMBH masses correlate with both the X-ray temperature and luminosity, with the former exhibiting a better correlation (the correlation coefficient $r_{\rm P}\simeq0.73$ for $M_{\rm BH}-T_{\rm X}$ compared to $r_{\rm P}\simeq0.63$ for $M_{\rm BH}-L_{\rm X}$). For star-forming galaxies, we find statistically-significant but weaker correlations between SMBH masses and X-ray quantities, with $r_{\rm P}\simeq0.52$ for $M_{\rm BH}-T_{\rm X}$ and $r_{\rm P}\simeq 0.41$ for $M_{\rm BH}-L_{\rm X}$.

The gas mass appears to be weakly correlated with the SMBH mass even for the quenched population of galaxies ($r_{\rm P}\simeq0.54$), suggesting that something else rather than SMBH mass (e.g. SMBH feedback) is more closely connected to (or even determines) the gas mass content within galaxies, as it has in fact been demonstrated in the case of the Illustris/TNG models and their variations \citep{zinger.etal.2020}, and all the way to halo scales \citep{genel.etal.2015, pillepich.etal.2018a} -- see also Section~\ref{sec:theo}. In the same vein, the weaker correlations between SMBH mass and X-ray gas properties of star-forming galaxies suggest that the latter may be more affected by non-gravitational processes, e.g. stellar and SMBH feedback, than is the case for quenched galaxies.  Finally, the emission-weighted metallicity appears to be slightly negatively correlated with the SMBH mass. 

These results imply that the correlation between the SMBH mass and the X-ray luminosity may be mostly driven by the correlations of the X-ray temperature as well as the halo mass with SMBH mass (as suggested by Eq.~\ref{eqn:5}). While in the high temperature regime ($T_{\rm X}>1$ keV) the X-ray emission can be well described by thermal Bremtrahlung, $L_{\rm X}\propto T^{1/2}$, at the low-temperature end ($T_{\rm X}<1$ keV) line emission contributes significantly to the total X-ray emission. This might partly account for the weaker  $M_{\rm BH}-L_{\rm X}$ correlation compared to the $M_{\rm BH}-T_{\rm X}$ relation.

%%%%%%%%%%%%%%%%%%%%%%%% Table %%%%%%%%%%%%%%%%%%%%%%%
 \begin{table*}
  \caption{\label{tb1}
  Best-fitting parameters for the SMBH--X-ray scaling relations, $M_{\rm BH}-T_{\rm X}$ and $M_{\rm BH}-L_{\rm X}$, as presented in the form of Eq.~\ref{eqn:10} for the TNG100 quenched galaxies, separately for all galaxies and for central galaxies only. In addition we also provide values for correlation coefficients ($r_P$). For comparison, we report the corresponding results from observations, where available.}
 \begin{center}
  \resizebox{0.99\textwidth}{!}{
 \begin{tabular}{c|ccccc|c|ccccc}
 \hline
   &  &  & $M_{\rm BH}-T_{\rm X}$   &  & & $\mid$ & & & $M_{\rm BH}-L_{\rm X}$  & \\
 \hline
Relation & $F_0$ & Normalization & Slope & Scatter & $r_P$ & $\mid$ & $F_0$ & Normalization & Slope & Scatter & $r_P$ \\
\hline
 {\bf Within $R_{\rm e}$}  \\
 TNG100 Central+Satellite & 1 keV & $1.11\pm0.02$ & $1.06\pm0.03$ & $0.26\pm0.01$ & 0.76 & $\mid$ & $10^{44}$ erg/s & $1.89\pm0.06$ & $0.25\pm0.01$ & $0.30\pm0.01$ & 0.63 \\
 TNG100 Central only & 1 keV & $1.12\pm0.02$ & $1.07\pm0.04$ & $0.25\pm0.01$ & 0.79 &$\mid$ & $10^{44}$ erg/s& $1.97\pm0.06$ & $0.27\pm0.01$ & $0.29\pm0.01$ & 0.68\\
\cite{lakhchaura.etal.2019} BCG & 1 keV & $1.50\pm0.08$ & $2.30\pm0.40$ & $0.26\pm0.08$ & $0.88$ & $\mid$ &$10^{44}$ erg/s & $2.34$ & $0.41\pm0.13$ & $0.42\pm0.11$ & 0.66   \\
\cite{gaspari.etal.2019} & $1$ keV & $1.39\pm0.05$ & $2.70\pm0.17$ & $0.21\pm0.03$ & 0.94 &$\mid$ &$10^{44}$ erg/s & $2.63\pm0.15$ & $0.51\pm0.04$ & $0.30\pm0.02$ & $0.87$ \\
 \hline
 {\bf Within $5R_{\rm e}$}  \\
 TNG100 Central+Satellite & 1 keV & $1.18\pm0.01$ & $1.23\pm0.03$ & $0.20\pm0.01$ & 0.86 &$\mid$ & $10^{44}$ erg/s & $1.76\pm0.04$ & $0.27\pm0.01$ & $0.28\pm0.01$ & 0.69 \\
 TNG100 Central only & 1 keV & $1.19\pm0.02$ & $1.23\pm0.03$ & $0.19\pm0.01$ & 0.88 &$\mid$ & $10^{44}$ erg/s & $1.77\pm0.05$ & $0.27\pm0.01$ & $0.28\pm0.01$ & 0.72 \\
 \cite{lakhchaura.etal.2019} BCG & $1$ keV & $1.37\pm0.08$ & $1.90\pm0.34$ & $0.25\pm0.08$ & $0.88$ & $\mid$ &$10^{44}$ erg/s & $2.08$ & $0.37\pm0.12$ & $0.40\pm0.11$ & 0.70  \\
 \hline
 {\bf Within $R_{\rm 500c}$}  \\

TNG100 Central only & 1 keV & $1.35\pm0.02$ & $1.44\pm0.03$ & $0.19\pm0.01$ & 0.89 & $\mid$ & $10^{44}$ erg/s & $1.68\pm0.03$ & $0.28\pm0.01$ & $0.22\pm0.01$ & 0.83\\
\cite{bogdan.etal.2018} & $1$ keV & $1.20\pm0.09$ & $1.74\pm0.16$ & $0.38$ & $0.97$ & $\mid$ \\
\cite{gaspari.etal.2019} & $1$ keV & $1.18\pm0.05$ & $2.14\pm0.13$ & $0.25\pm0.02$ & $0.93$ & $\mid$ & $10^{44}$ erg/s & $2.00\pm0.11$ & $0.38\pm0.3$ & $0.31\pm0.02$ & $0.86$ \\
 \hline
 \end{tabular}}

 \end{center}
 \end{table*}
 %%%%%%%%%%%%%%%%%%%%%%%%%%%%%%%%%%%%%%%%%%%%%%%%%%%%%%%%%%%%%%%%%%%%%%%%%%%%5
\begin{figure*}
    \centering
    \includegraphics[width=0.9999\textwidth]{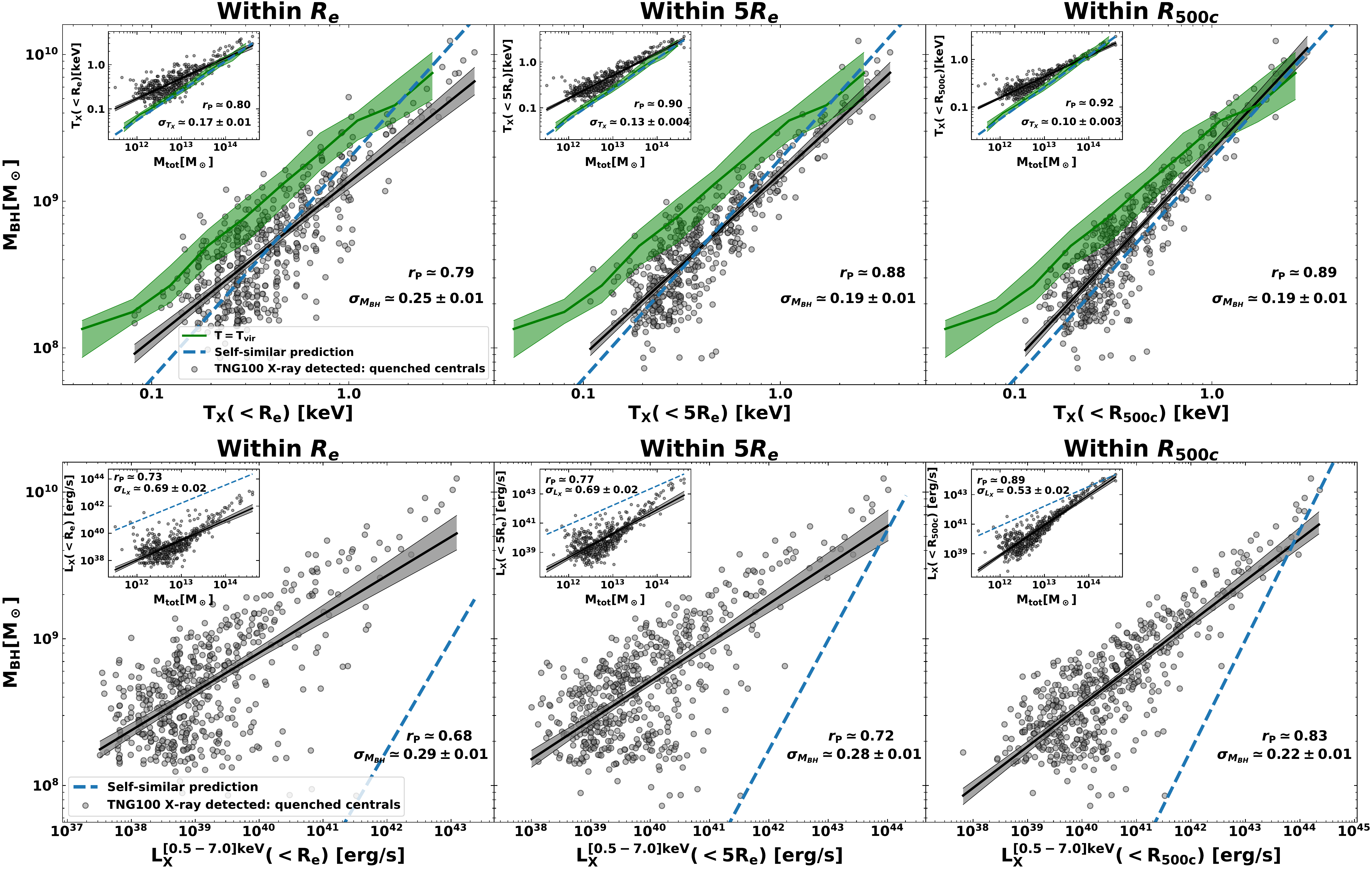}
    \caption{Inspection of the SMBH--X-ray scaling relations across different radial ranges over which the gas properties are integrated for a subsample of TNG100 quenched and central galaxies. The main panels show $M_{\rm BH}-T_{\rm X}$ (top row) and $M_{\rm BH}-L_{\rm X}$ (bottom row) relations at various radii, while the sub-panels represent the relations of the corresponding X-ray quantity with the total halo mass ($M_{\rm tot}$). The green lines and shaded areas represent the median relations and $1\sigma$ uncertainties where $T_{\rm X}=T_{\rm vir}$.The normalisation of the self-similar predicted relations (dashed lines) is determined by assuming that the X-ray quantities measured within the largest considered radius, i.e within $R_{\rm 500c}$, of the most massive galaxies obey the self-similar prediction (see Section~\ref{sec:self-similar}). For each scaling relation, the Pearson coefficient ($r_{\rm P}$) as well as the intrinsic scatter ($\sigma$, in dex) are provided.}  
    \label{fig:e4}
\end{figure*}

%%%%%%%%%%%%%%%%%%%%%%%%%%%%%%%%%%%%%%%%%%%%%%%%%%%%%%%%%%%%%%%%%%%%%%%%%
\subsection{Radial dependence of the SMBH--gas atmospheres correlations for the quenched population}
\label{sec:radial}
%%%%%%%%%%%%%%%%%%%%%%%%%%%%%%%%%%%%%%%%%%%%%%%%%%%%%%%%%%%%%%%%%%%%%%%%%

We now inspect the correlations between the masses of central SMBHs and the X-ray properties of the hot atmospheres as a function of radius, i.e. of galactocentric distance within which the X-ray signals are integrated (see also Section~\ref{sec:method}). 
In light of the results presented in the previous Sections, where we find that quenched galaxies exhibit the most significant SMBH-X-ray scaling relations, we here focus on quenched galaxies and investigate the physical origin of such strong correlations.

In Fig.~\ref{fig:e4} (main panels and gray points and curves), we present the relation between the central SMBH mass and the X-ray temperature (top) and luminosity (bottom) of the gas atmospheres of TNG100 quenched host galaxies that could be detected with 100ks exposure with {\it Chandra}. The different columns depict results at three different projected radii: within the galactic bodies ($<R_{\rm e}$), within the gaseous coronae ($<5R_{\rm e}$), and throughout the intragroup/cluster media ($<R_{\rm 500c}$), from left to right. 
The best-fitting parameters, i.e. normalization, slope, as well as intrinsic scatter, are reported in Table~\ref{tb1}. 

In addition to these relations, in the insets of Fig.~\ref{fig:e4}, we also show the scaling relations of the X-ray gas properties with the total halo mass, i.e. $T_{\rm X}-M_{\rm tot}$ and $L_{\rm X}-M_{\rm tot}$. 

In TNG, the X-ray temperatures appear to correlate with SMBH masses better than the X-ray luminosities, not only within the galactic regions as shown in the previous Section, but across the entire radial range. Importantly, we notice a radial trend whereby the SMBH scaling relations with X-ray gas quantities become stronger, i.e. with higher $r_{\rm P}$, and tighter, i.e. with smaller intrinsic scatter ($\sigma_{M_{\rm BH}}$), towards larger radii. 

The scaling relations $M_{\rm BH}-T_{\rm X}$ and $M_{\rm BH}-L_{\rm X}$ within $R_{\rm 500c}$ display the most significant correlation between SMBH masses and X-ray properties of the gas, with correlation coefficient reading $0.89$ ($0.83$) for $T_{\rm X}$ ($L_{\rm X}$). Crucially, for intermediate or larger radii ($r\gtrsim5R_{\rm e}$, i.e. for gas properties integrated or mediated within the circum-galactic or intra-group/cluster medium), the $M_{\rm BH}-T_{\rm X}$ scaling relation of TNG quenched galaxies displays remarkably small intrinsic scatter with $\sigma_{M_{\rm BH}}\lesssim0.2$ dex. Inspecting the scaling relations between the total halo mass and the X-ray properties shown in the sub-panels of Fig.~\ref{fig:e4}, the $T_{\rm X}-M_{\rm tot}$ and $L_{\rm X}-M_{\rm tot}$ relations exhibit similar radial trends as the SMBH--X-ray scaling relations, in terms of both their strength and tightness.
Analog plots of those in Fig.~\ref{fig:e4} but focused on the central star-forming galaxies in TNG100 (not shown) reveal, on the other hand, less significant trends of the correlations between SMBH mass and X-ray gas quantities as a function of aperture; they also exhibit overall weaker correlations ($r_{\rm P}\sim0.4-0.5$ throughout) than in the quenched case, in line with what already seen in the previous Sections.

Since gravity is the major player that defines the hot atmosphere's thermodynamics at large halo/spatial scales ($<R_{\rm 500c}$, at least according to the self-similar ansatz), the correlations and their radial trends between SMBH mass and X-ray gas properties in both the real Universe and in TNG could be due to an underlying relationship between the total halo mass and the SMBH mass, as described in Section~\ref{sec:bhgrowth} for TNG100. However, gravity is unlikely the sole player. We notice that for all the considered relations, a single power law appears to fit reasonably well the simulated data at large radii (i.e. within $R_{\rm 500c}$), while within smaller apertures, there are offsets between the best-fit relations and the data points of the most massive objects, as clearly visible, for instance, in the case of the X-ray luminosity relations. This result hints at a relatively stronger influence on gas at progressively smaller galactocentric distances and for lower-mass galaxies of other astrophysical processes beyond gravity, such as SMBH feedback -- which in TNG is the main non-gravitational channel for energy-injection in the considered mass range (\citealt{weinberger.etal.2018}). 

%This result may suggest that the X-ray temperatures of hot atmospheres at large radii are good proxies of their host galaxy mass thereby reflecting reasonably well the underlying relationship between galaxy total mass and its central SMBH mass, as described in Section~\ref{sec:bhgrowth}.

% not necessary to draw any conclusion here, just state the direction that gravity appears to define the X-ray scaling relations at high-mass end and at large radii, while the scaling relations are more susceptible to presumably SMBH feedback at low-mass end and at small radii.
%%%%%%%%%%%%%%%%%%%%%%%%%%%%%%

\subsection{Comparison to the self-similar predictions}
%%%%%%%%%%%%%%%%%%%%%%%%%%%%%%%%%

To further understand the interplay of gravity and SMBH feedback in shaping the SMBH -- X-ray gas scaling relations, we examine how well such relations from the TNG quenched population reflect self-similarity. 

In Fig.~\ref{fig:e4}, the self-similar predictions as obtained in Section~\ref{sec:self-similar} are represented by dashed blue lines. The normalisation of Eqs.~\ref{eqn:8})-(\ref{eqn:9} are determined by assuming that the X-ray measurements within the largest radii ($R_{\rm 500c}$) of the most massive objects satisfy the self-similar predictions.

Firstly, the blue dashed lines in Fig.~\ref{fig:e4} are meant to visualize what we have already emphasized in Section~\ref{sec:self-similar}: correlations between halo gas properties and total halo mass are in place also in the absence of SMBHs and their feedback. Therefore, a priori, the existence of relations between SMBH masses and gas properties could emerge by assuming that SMBH growth is connected to total halo growth, without necessarily requiring any causal physical connections between SMBHs and gas atmospheres -- such as those that could be mediated by SMBH feedback -- and without invoking any dependence of SMBH growth on the availability of accretion gas in the innermost regions of galaxies.

On the other hand, the insets of Fig.~\ref{fig:e4} demonstrate that, even in the presence of complex astrophysical processes such stellar and SMBH feedback, as in general is the case in TNG, in simulations like TNG, strong and tight correlations between X-ray gas properties and total halo mass naturally exist around quenched galaxies {\it also for the hot gas at small galactocentric distances} (e.g. within $R_{\rm e}$ or $<5R_{\rm e}$), albeit somewhat weaker and less tight than in the halo-scale regime (left versus right columns). However, importantly, the shapes or slopes of such relations (and those between SMBH masses and X-ray gas properties) do indeed differ in TNG100 from the self-similar (+ merger-only driven) predictions of Section~\ref{sec:self-similar}, and more so for the gas closer to the galaxies' centers and for lower-mass galaxies.

For the gas X-ray temperature, the TNG-predicted relations for quenched galaxies are shallower than the self-similar expectations at all scales, but for the $M_{\rm BH}-T_{\rm X}$ correlations at the halo scales (i.e. $<R_{\rm 500c}$, top right panel).
In fact, when comparing the TNG $M_{\rm BH}-T_{\rm X}$ relation with the self-similar predictions, one has to take into account the comparison between the TNG $M_{\rm BH}-M_{\rm tot}$ relation and the self-similar predicted connections (i.e. those merger-driven) presented in Section~\ref{sec:bhgrowth}. We note that SMBH mass and atmospheric temperature in TNG, at a given halo mass, deviate from the self-similar prediction in a similar manner, namely, the two simulated quantities are larger than the self-similar values and the deviation is more pronounced towards the low-mass end. Interestingly, because of the similar deviations of the $M_{\rm BH}-M_{\rm tot}$ and $T_{\rm X}-M_{\rm tot}$ relations from self-similarity, the resulting $M_{\rm BH}-T_{\rm X}$, particularly at $R_{\rm 500c}$, is statistically in line with the self-similar prediction (with slope of about 1.5). 

Another way to compare to gravity-only driven expectations is by comparing the effective X-ray temperatures of the gas atmospheres from the simulations to the virial temperature, i.e. $T_{\rm vir}$, as predicted by the self-similar model (represented by green lines in Fig.~\ref{fig:e4}\footnote{In fact, in this plots, the virial temperature is determined via $M_{200c}$ (see Eq.~\ref{eq_tvir}) rather than the exact virial mass ($M_{\rm vir}$), hence the scatter.}). 
%
%In this way, we observe that the $M_{\rm BH}-T_{\rm vir}$ relation is always shallower than the self-similar relation as it only captures the self-similar deviation of the SMBH mass.
%
The TNG simulations produce hotter gaseous atmospheres with respect to the virial temperature at any given total halo mass (insets), and the offset appears to be more prominent at smaller radii and smaller masses, resulting in a shallower slope for the simulated $T_{\rm X}-M_{\rm tot}$ relations ($\sim0.44-0.49$) than the self-similar and virial temperature cases ($\sim0.67$), across the three considered apertures. On the other hand, the $M_{\rm BH}-T_{\rm vir}$ relation is typically shallower than the self-similar relation and the TNG outcome, reflecting mostly the deviation in slope in TNG100 from the self-similar, and merger-only driven , $M_{\rm BH}-M_{\rm tot}$ relation (see Fig.~\ref{fig:1}).
%as it only captures the deviation from self-similarity of the SMBH mass. AP: I would not phrase it like that, Tvir does know nothing about BHs)

In contrast to the gas temperature, the TNG X-ray luminosity of quenched galaxies at a given halo mass (insets in the lower row of Fig.~\ref{fig:e4}) is largely underestimated in comparison to the self-similar values across all considered radial apertures and particularly in the lower-mass systems. Furthermore, the simulated $L_{\rm X}-M_{\rm tot}$ relations become steeper than the self-similar ones for gas properties measured on larger apertures.
Unlike the other quantities such gas temperature and SMBH mass, the X-ray luminosities in TNG deviate from the self-similar prediction in the opposite direction. The contrary self-similar deviations of the $M_{\rm BH}-M_{\rm tot}$ and $L_{\rm X}-M_{\rm tot}$ relations with respect to the gravity+merger only driven scenarios collectively result in significantly shallower $M_{\rm BH}-L_{\rm X}$ relations in TNG100 quenched galaxies compared to the self-similar one, at all radii.

In the case of star-forming galaxies, the TNG100 predictions (not shown) deviate even more strongly from the self-similar expectations than those depicted in Fig.~\ref{fig:e4} for quenched galaxies.

The deviations of the TNG SMBH -- X-ray gas scaling relations from self-similarity can be explained by SMBH feedback and by the dependence of SMBH growth on the availability of gas, particularly for the lower-mass systems with total halo mass $\lesssim 10^{11-12}~M_{\rm tot}$. It is the gas accretion onto SMBHs, as discussed in Section~\ref{sec:bhgrowth}, that enhances their mass relative to the predicted linear, merger-only driven, growth. The gas accretion also triggers SMBH feedback which, in turn, affects the thermodynamic and density properties of the gaseous atmospheres. In the case of massive, mostly quenched galaxies, such feedback not only heats the gaseous atmospheres but also expels a large amount of the gas content out of the central regions of galaxies and haloes (\citealt{davies.etal.2019,zinger.etal.2020}). Consequently, the preventative and ejective effects of SMBH feedback are the main driver of the increase of the gas temperature as well as the reduction of X-ray luminosity with respect to the self-similar predictions. 

%%%%%%%%%%%%%%%%%%%%%%%%%%%%%%%%%%%%%%%%%%%%%%%%%%%%%%%%%%%%%%%%%%%%%%%%%%%%%%%%%
\begin{figure*}
    \centering
    \includegraphics[width=0.95\textwidth]{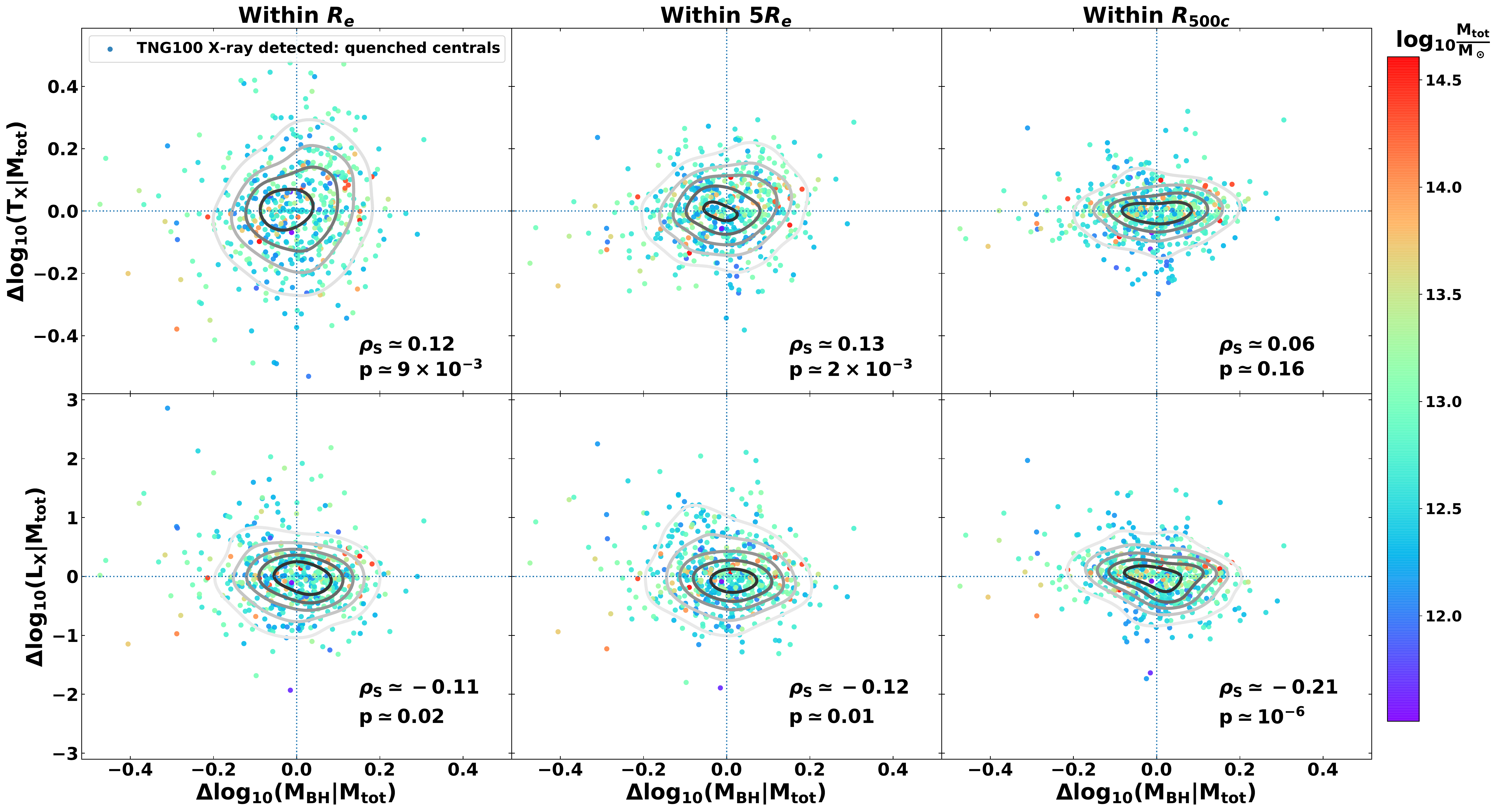}
    \includegraphics[width=0.95\textwidth]{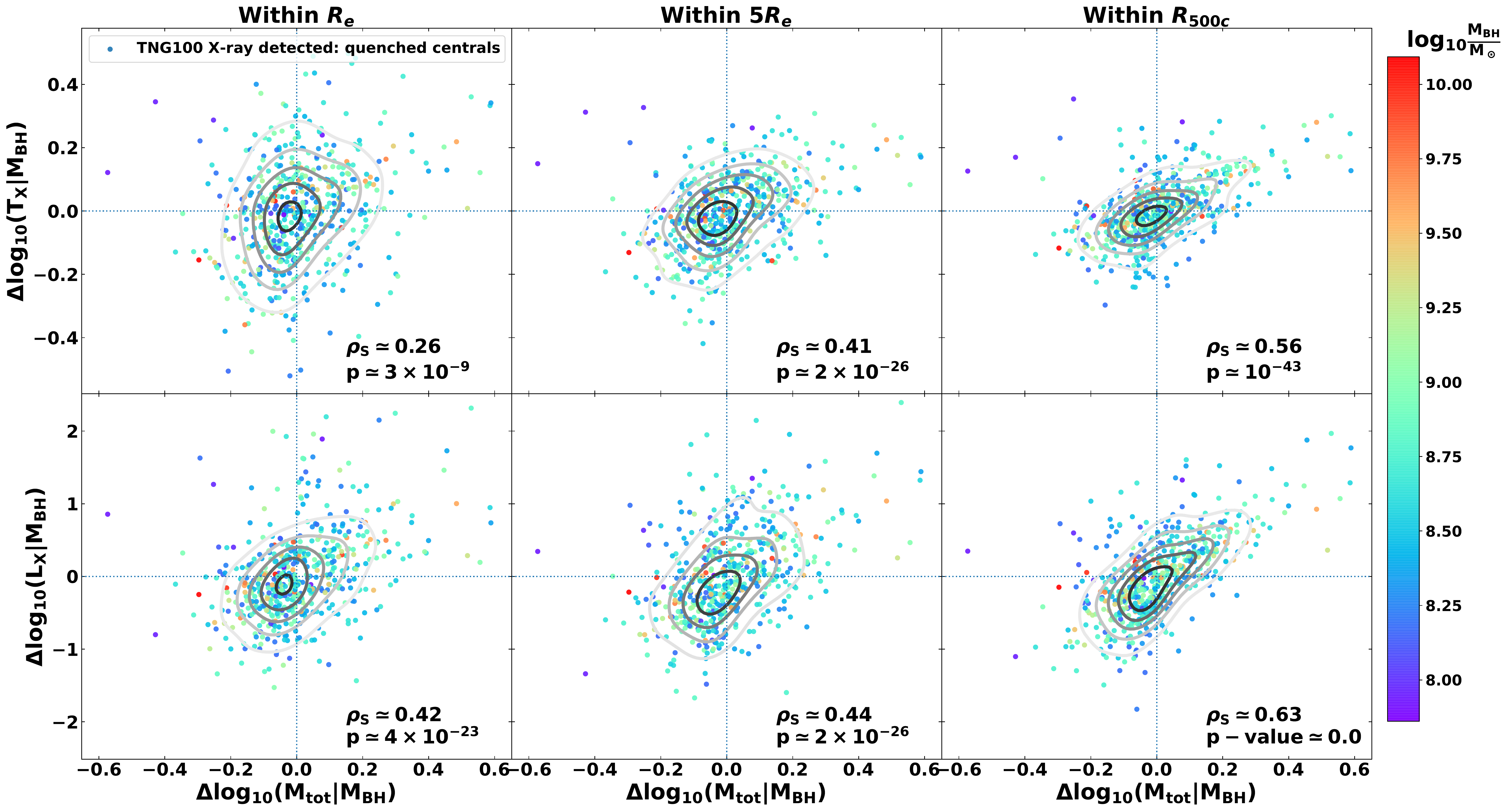}
    \caption{{\it Top panel}: Correlation between variations in X-ray quantities, $\Delta T_{\rm X}$ (upper row) and $\Delta L_{\rm X}$ (lower row), and variation in SMBH mass ($\Delta M_{\rm BH}$) at a given total halo mass ($M_{\rm tot}$), for a subsample of TNG100 quenched and central galaxies at $z=0$. The variations are computed with respect to the corresponding median value at a given mass bin. To quantify the correlation between those variations, Spearman correlation coefficients ($\rho_{\rm S}$) are provided as well as the corresponding p-value. {\it Bottom panel:} similar to the top panel but the correlation is now between variations in X-ray quantities and variation in total halo mass ($\Delta M_{\rm tot}$) at a fixed SMBH mass ($M_{\rm BH}$).}  
    \label{fig:e5}
\end{figure*}
%%%%%%%%%%%%%%%%%%%%%%%%%%%%%%%%%%%%%%%%%%%%%%%%%%%%%%%%%%%%%%%%%%%%%%%%%%%%%%%%%

%%%%%%%%%%%%%%%%%%%%%%%%%%%%%%%
\subsection{Impact of SMBH feedback versus halo mass dependence}
%%%%%%%%%%%%%%%%%%%%%%%%%%%%%%%
 Is it possible to isolate and quantify the influence of SMBH feedback on the X-ray properties of the hot gaseous atmospheres from their dependence on the total halo mass, a dependence which is driven by gravity and by the hierarchical growth of structure? We attempt to do so next. 
 
 For this task, we examine the correlation between variations in X-ray quantities, i.e. $\Delta T_{\rm X}$ and $\Delta L_{\rm X}$, and variation in SMBH mass ($\Delta M_{\rm BH}$), where the variations are estimated with respect to the corresponding median values at the same total halo mass bins (upper half of Fig.~\ref{fig:e5}). For comparison, we also show the same exercise for the correlations between X-ray variations and total halo mass variations ($\Delta M_{\rm tot}$) at the same SMBH mass bins (lower half). 
 
 The inspection of the correlations across the usual considered radial ranges is shown in Fig.~\ref{fig:e5}, in which Spearman correlation coefficients ($\rho_{\rm S}$) as well as the corresponding p-values are provided in order to quantify the strength and the significance of the correlations.
 
 As shown in the top panel of Fig.~\ref{fig:e5}, there are overall very weak positive (negative) correlations between $\Delta T_{\rm X}$ ($\Delta L_{\rm X}$) and $\Delta M_{\rm BH}$ across the considered ranges, demonstrated by low values of the Spearman coefficients ($|\rho_{\rm S}|\lesssim0.2$), and the correlations are also statistically insignificant in most of the considered apertures where the p-values are relatively large ($p>10^{-3}$). Notably, the variation in X-ray luminosity is, although weak, negatively correlated with the variation in SMBH mass. This could be the result of the ejective effect of SMBH feedback in low-mass galaxies, as discussed in the previous Section. 
 
 On the other hand, the bottom panel of Fig.~\ref{fig:e5} shows that there are stronger ($\rho_{\rm S}\simeq0.2-0.6$) and more significant ($p\lesssim10^{-9}$) correlations between variations in X-ray quantities and variations in total halo mass considered at the same SMBH mass bins. Particularly toward larger radii, both variations in temperature and X-ray luminosity become to be more positively correlated with the total mass variations with $\rho_{\rm S}>0.5$. \\

 \subsection{Summary on correlations and possible physical drivers}
 In summary, the results presented in this Section~\ref{sec:3} support a picture whereby, according to the TNG model, the X-ray properties of the hot gaseous atmospheres -- particularly in quenched galaxies ($M_{\rm BH}\gtrsim10^{8.2}M_\odot$) -- are, to a first approximation and also at relative small galactocentric distances (i.e. within a few stellar effective radii), determined by the gravitational potential of the host haloes. On top of that, SMBH feedback, via ejective and preventative effects, causes the X-ray properties to deviate from the self-similar predictions. These secondary, albeit possibly very informative, effects appear to be relatively more prominent toward low-mass and small-scale regimes, i.e. at the group mass scales and below, and for the circum-galactic gas closer to the galaxies' centers. In parallel, strong and tight correlations between the mass of SMBHs and the X-ray properties of the gaseous atmospheres are a natural outcome of state-of-the-art galaxy formation models like IllustrisTNG, particularly so for quiescent galaxies and when the gas properties are integrated or mediated over halo, i.e. large, spatial scales. However, to a first order approximation, the very existence of relations between SMBHs and gas atmospheres seems to us to be mainly a reflection of the underlying $M_{\rm BH}-M_{\rm tot}$ relations. 
Within this interpretation framework, the weaker correlations between SMBH mass and X-ray gas properties found for the star-forming galaxies (Section~\ref{sec:BH-Xray_corr}) could also be a reflection of a weaker connection between SMBH mass and total halo mass: while the latter relation holds also for star-forming galaxies in TNG, it is weaker than for quenched galaxies ($r_{\rm P}\sim0.6$ vs. $r_{\rm P}\sim0.85$), and this could at least partly explain why the $M_{\rm BH}-T_{\rm X}\ (L_{\rm X})$ correlations appear to be stronger for the quenched than for the star-forming galaxies. Overall, these comparisons suggest that the X-ray observables of star-forming galaxies may actually encode, contrary to expectations, more direct information about the functioning of SMBH feedback than those of quenched galaxies. 

%%%%%%%%%%%%%%%%%%%%%%%%%%%%%%%%%%%%%%%%%%%%%%%%%%%%%%%%%%%%%%%%%%%%%%%

%%%%%%%%%%%%%%%%%%%%%%%%%%%%%%%%%%%% Fig%%%%%%%%%%%%%%%%%%%%%%%%%%%%%
\begin{figure*}
    \centering
   \includegraphics[width=0.95\textwidth]{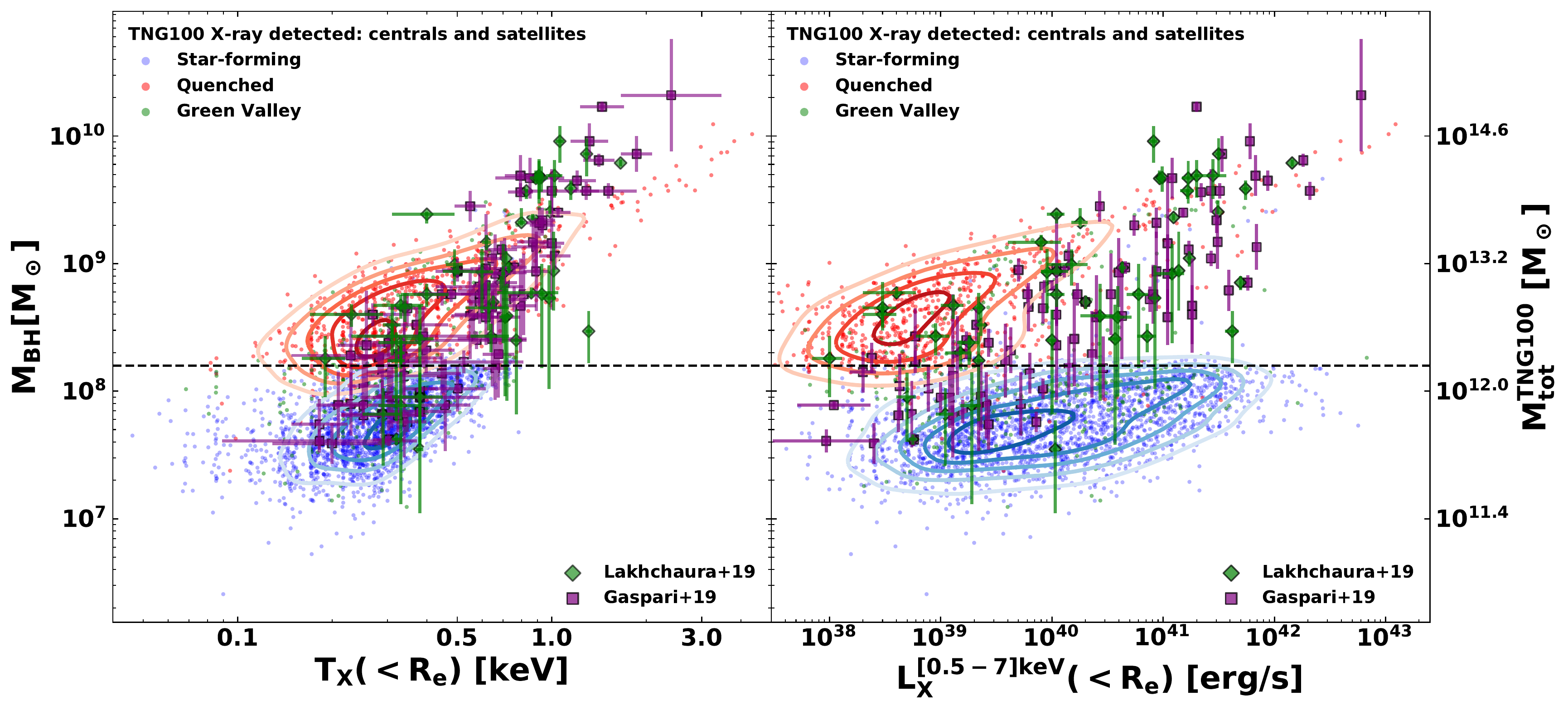}
    \includegraphics[width=0.95\textwidth]{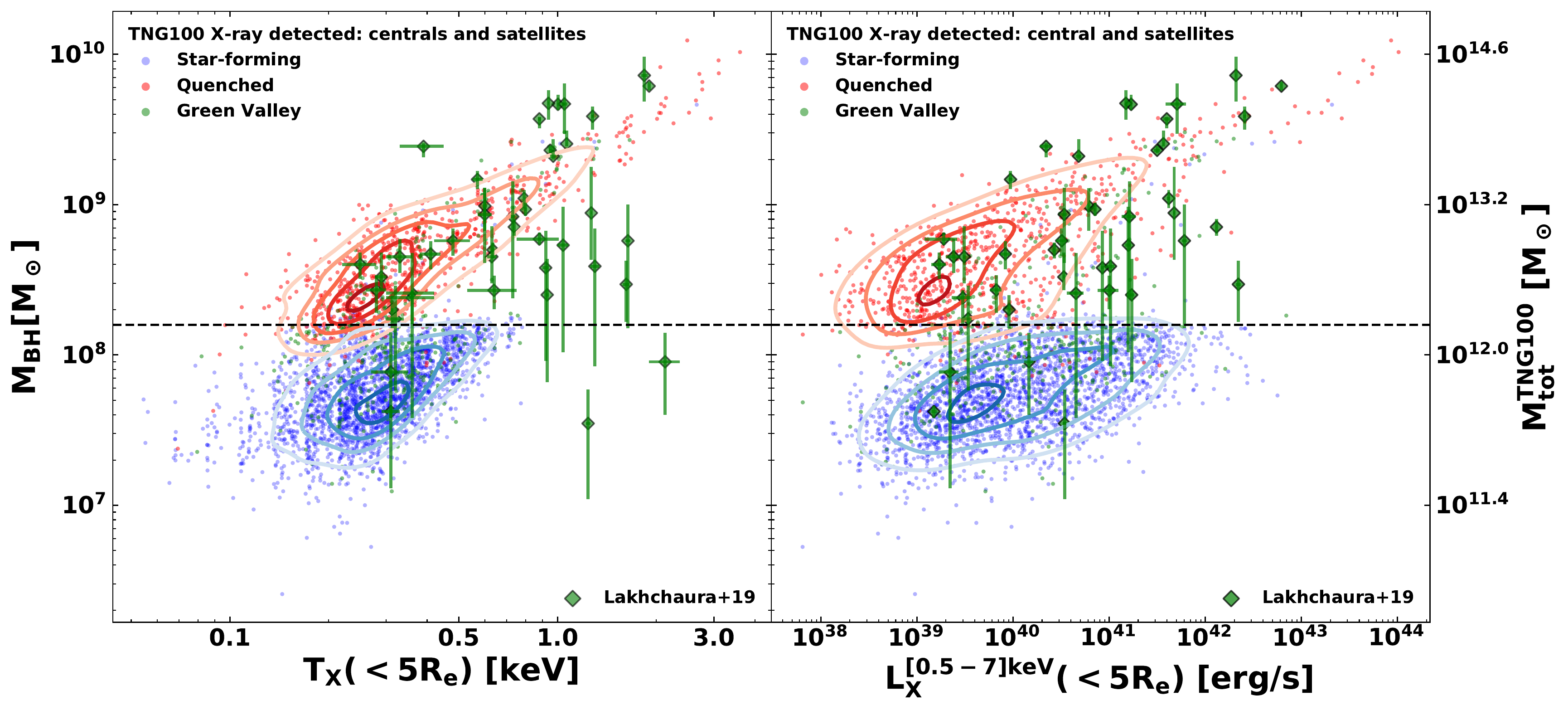}
    \includegraphics[width=0.95\textwidth]{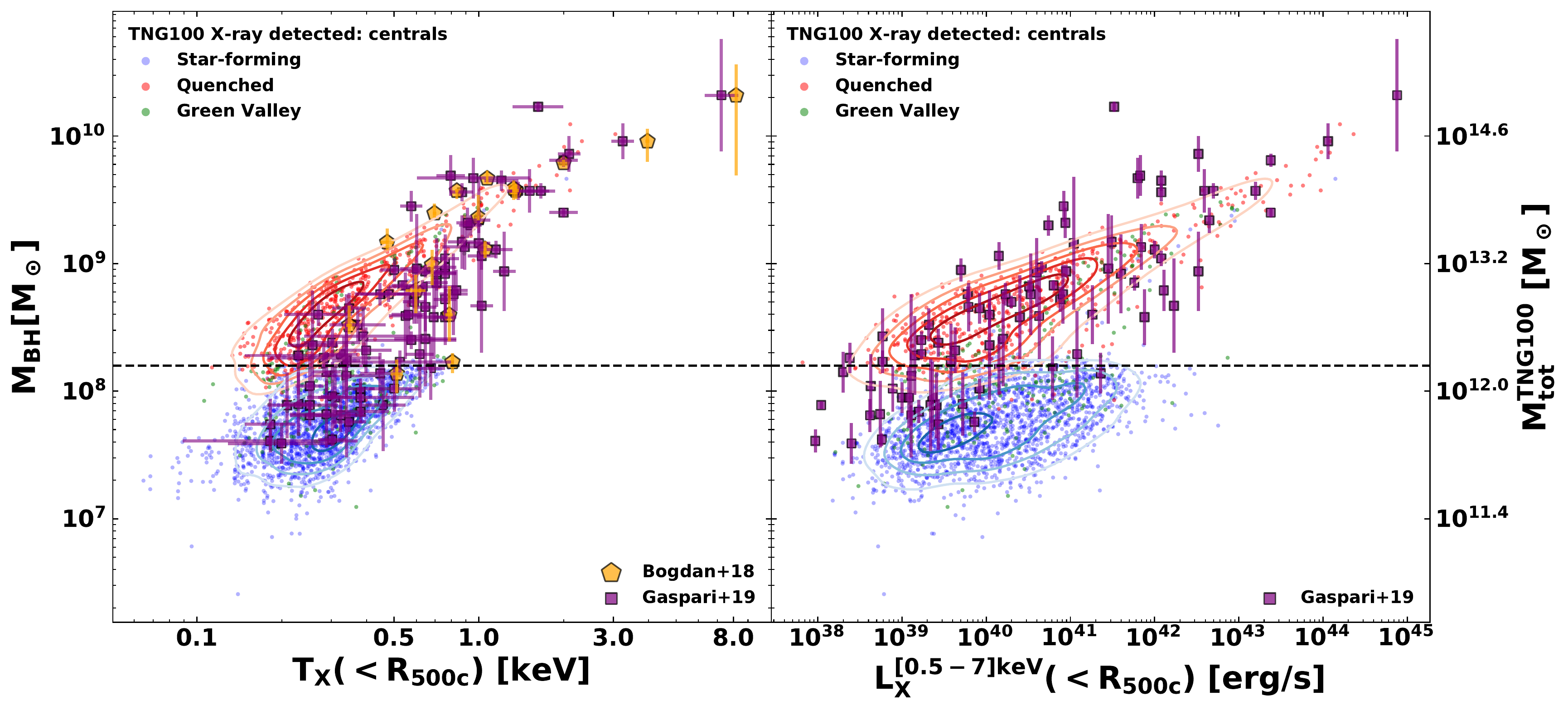}
    \caption{Comparison between the TNG100 simulation and X-ray observations on the SMBH--hot atmospheres correlations at various apertures, from top to bottom. The red and blue contours specify the loci of quenched and star-forming TNG100 populations, respectively, including only simulated galaxies that could be detected with 100 ks exposure with {\it Chandra}. The horizontal dashed line marks $M_{\rm BH}\sim10^{8.2}M_\odot$ where is approximately the boundary line between quenched and star-forming populations. For reference, the rightmost y-axis specifies the median value of TNG100 $M_{\rm tot}$ according to the median relation $M_{\rm BH}-M_{\rm tot}$ as shown in Fig.~\ref{fig:1}.} 
    
    \label{fig:3}
\end{figure*}

%%%%%%%%%%%%%%%%%%%%%%%%%%%%%%%%%%%%%%%%%%%%%%%%%%%%%%%%%%%%%%%%%%%%%%%%%
\section{Comparison with X-ray Observations}
\label{sec:comparison_xray}
%%%%%%%%%%%%%%%%%%%%%%%%%%%%%%%%%%%%%%%%%%%%%%%%%%%%%%%%%%%%%%%%%%%%%%%%%%

To gauge to what level we can apply the insights gained from IllustrisTNG to the real Universe, we here compare the relations from the simulations to X-ray observations of nearby galaxies taken from \cite{bogdan.etal.2018}, \cite{lakhchaura.etal.2019}, and \cite{gaspari.etal.2019}, as described in Section~\ref{sec:xraydata}.
%It is worth mentioning a caveat in comparing simulations with observations: as the observational datasets are not well-defined and incomplete, it prevents us from adopting a more tailored approach in selecting simulated systems besides the fact that we select only quenched simulated galaxies. 

In Fig. \ref{fig:3} we show the TNG100 simulation outcome overlaid to those from observations for the SMBH scaling relations at three different radii: within $R_{\rm e}$, $5R_{\rm e}$, and $R_{\rm 500c}$, from top to bottom. For the comparison, we employ the full X-ray detected sample of TNG100 at z=0 by including both star-forming and quenched galaxies. The reason is twofold: on the one hand, because the selection functions of the observed samples are not well-defined or known, we cannot construct galaxy samples from the simulation that correspond in detail to the observed ones; on the other hand, we wish to more fairly and broadly compare with mixed-type observational samples, e.g. the one obtained by \cite{gaspari.etal.2019}. For reference, we also report best-fit parameters for the observed relations, given in the form of Eq.~\ref{eqn:10}, in Table~\ref{tb1}. Nonetheless and before proceeding, we wish to stress that, as the current observed samples are far from being complete as well as they are not selected based on homogeneous and well-defined criteria, we should not over-interpret, quantitatively, the comparison between simulated and observed parameters and data points provided in Table~\ref{tb1} and Fig. \ref{fig:3}, respectively.   

From Fig. \ref{fig:3}, we can deduce that the TNG simulation predictions qualitatively align with recent X-ray observations, which also reveal remarkably strong and tight correlations between SMBH masses and the hot X-ray emitting atmospheres, from galactic to cluster scales. As reproduced in Fig. \ref{fig:3}, \cite{lakhchaura.etal.2019} found that there is a strong correlation ($r_{\rm P}\sim0.88$) between $k_{\rm B}T_{\rm X}$ (within $R_{\rm e}$ and $5R_{\rm e}$) and $M_{\rm BH}$ for a subsample of 18 BCGs with intrinsic scatter of $\sim0.26$ dex. Similarly, \cite{gaspari.etal.2019} showed that for a sample of 85 galaxies of various types the atmospheric gas temperature measured within the galactic/circum-galactic scale and within the group/cluster core does strongly correlated with the SMBH mass with $r_{\rm P}>0.9$. The intrinsic scatter of those relations is relatively small $\sim0.2-0.25$ dex. At cluster/halo scales ($<R_{\rm 500c}$, bottom panels), an X-ray study from \cite{bogdan.etal.2018} found a strong correlation between the ICM temperature and the central SMBHs, with $r_{\rm P}\sim0.97$, though with a large scatter ($\sim0.4$ dex). 

Regarding the radial trends found in TNG, the analysis of the relatively small sample of BCGs from \cite{lakhchaura.etal.2019} indicates that the scatter in the $M_{\rm BH}-T_{\rm X}$ relation ($\sim0.26$ dex) does not change significantly between $R_{\rm e}$ and $5R_{\rm e}$. On the other hand, the data from \citealt{gaspari.etal.2019} suggest that the $M_{\rm BH}-T_{\rm X}$ relation at cluster core scale ($\sim0.1R_{\rm 500c}$) has larger intrinsic scatter compared to the one within a smaller galactic/CGM scale, $0.25\pm0.02$ versus $0.21\pm0.03$, i.e. opposite to the simulation trend. However, it is worth mentioning that the radial ranges considered in \cite{gaspari.etal.2019} are not homogeneous, e.g. the galactic/CGM range consists of $1-3R_{\rm e}$.

Regarding the intrinsic scatter, the simulation's X-ray scaling relations appear to have heterogeneous scatter across the considered dynamical range, with the scatter becoming smaller at higher temperatures (or SMBH masses). For instance, TNG100 galaxies with $kT_{\rm X}\lesssim1$ keV appear to exhibit $\sim10\%$ greater scatter with respect to the best-fit relation in comparison to those at higher temperatures. This result is in line with \citealt{lakhchaura.etal.2019} who found that the X-ray correlations are only significant for the massive BCG subsample, while for the less massive non-BCGs and lenticular galaxies \citealt{lakhchaura.etal.2019} find no significant correlation. On the other hand, the compiled dataset from \cite{gaspari.etal.2019} shows no visible variation for the scatter of X-ray scaling relations across the whole considered dynamical range. The sample of \cite{bogdan.etal.2018} is too small to carry out a meaningful assessment regarding the scatter of the $M_{\rm BH}-T_{\rm X}$ relation.

In general the simulated and observed data occupy similar regions of the parameter space with the level of consistency varying across the considered apertures.
\begin{itemize}
    \item Within $R_{\rm e}$: the simulated $M_{\rm BH}-T_{\rm X}$ relation tends to underpredict the SMBH masses of a few objects at the high-mass end, or conversely, to overpredict (underpredict) the gas temperatures at the high (low) mass end. It should be kept in mind that the observed data at this range display large uncertainties in both SMBH mass and X-ray gas temperature measurements. While acknowledging that TNG might produce discrepant X-ray properties for the hot atmospheres compared to X-ray observations (particularly within the ISM galactic regions, where the contribution to the X-ray signals from the star-forming gas cannot be assessed within the simulation model -- see Appendix C in \citealt{truong.etal.2020}), it is worth mentioning that estimating $T_{\rm X}$ within the central region of lower mass galaxies is challenging also observationally, due to the presence of unresolved X-ray point sources, which are accounted only in a statistical and model-dependent way (see e.g. \citealt{lakhchaura.etal.2018}). For the $M_{\rm BH}-L_{\rm X}$ relation, while the simulated and observed data points scatter over a similar region on the $M_{\rm BH}-L_{\rm X}$ plane, the TNG sample appears to be on average over-luminous (under-luminous) in comparison to the observed samples for systems with $M_{\rm BH}$ above (below) $\sim10^{8.2}M_\odot$. The discrepancy, which is about $1-2$ orders of magnitude in $L_{\rm X}$, could be in principle attributed to the nature of the TNG stellar and SMBH feedback models. Nonetheless, it is worth mentioning a factor that might partly contribute to the $L_{\rm X}$ discrepancy: the observed samples are incomplete and contain mixed-type galaxies which are not well classified, thereby it is not trivial to relate the observed luminosity, especially of those galaxies at the low-mass end, to either the simulated star-forming or the quenched population.
    
    \item Within $5R_{\rm e}$: TNG reproduces reasonably well the observed $M_{\rm BH}-T_{\rm X}$ and $M_{\rm BH}-L_{\rm X}$ relations, but there are some observed data points at the high-$T_{\rm X}$ end that appear to have higher $T_{\rm X}$ with respect to the simulated relation as well as the average observed trend. Those outliers are non-BCG galaxies belonging to clusters, such as NGC4473 and NGC4477 in the Virgo cluster, wherein the X-ray emission of those galactic atmospheres is likely to be contaminated by the emission from the ambient ICM. This could potentially explain the higher measured temperature of the hot gaseous atmospheres of these galaxies.   
    
    \item Within $R_{\rm 500c}$: both simulated TNG $M_{\rm BH}-T_{\rm X}$ and $M_{\rm BH}-L_{\rm X}$ relations are in good agreement with the observed data from \cite{bogdan.etal.2018} and \cite{gaspari.etal.2019}. We should however recall that here the integration boundary $R_{\rm 500c}$ for the simulation X-ray measurements is taken directly from the simulation output, it is the true one, i.e. it is not mocked.
\end{itemize}

A final caveat worth mentioning again for the interpretation of this comparison relates to what gas elements within simulated haloes are included in the X-ray mocks. In this Section, as throughout this paper, when selecting gas cells for the mock X-ray analysis, we employ the {\sc subfind} algorithm to select only gas cells that are gravitationally-bound to the central galaxies. This approach automatically excludes the contributions to the X-ray signals of gas parcels associated with satellite galaxies that belong to the same host halo, in line with the choices adopted in the highly-resolved observations of nearby galaxies reported in Fig. \ref{fig:3}. However, excising or not the X-ray contribution from satellite galaxies can have non negligible effects on the quantitative characterization of the SMBH--hot atmosphere correlations. This is explicitly demonstrated in Appendix~\ref{sec:app1}, where, for example, the contribution of satellite gas can strongly impact the X-ray gas temperature of low-mass systems ($<1$ keV) determined at large radii ($< R_{\rm 500c}$) and thus can change the slope and normalization of the derived $M_{\rm BH}-T_{\rm X}$ relation. In fact, according to TNG, including the gas contribution from satellite galaxies appears to jeopardise, i.e. worsen, the overall correlation between the X-ray gas temperatures and the central SMBH masses.
%%%%to-do list%%%%%%%%%%%%%%%
%%% provide the table of best-fitting parameters and updated the figure accrodingly. 

%%%%%%%%%%%%%%%%%%%%%%%% Table %%%%%%%%%%%%%%%%%%%%%%%
 \begin{table*}
  \caption{\label{tb2}
  Summary of the L25n512 model variations used for the study in Section~\ref{sec:theo}, performed on a box with the size of $\sim25\ {\rm Mpc/h}$ ($\sim37$ Mpc) .}
 \begin{center}
  \resizebox{0.99\textwidth}{!}{
 \begin{tabular}{lccc}
 \hline
Run name & Parameter(s) of interest & Fiducial value & Variation value\\
\hline
Fiducial & All parameters & See Table 1 in \cite{pillepich.etal.2018} & - \\
No Stellar Feedback & All stellar feedback related & See Table 1 in \cite{pillepich.etal.2018} & -\\
Stronger Stellar Feedback & Wind energy factor: $\overline{e}_w$ & 3.6 & 7.2 \\
Weaker Stellar Feedback & Wind energy factor: $\overline{e}_w$ & 3.6 & 1.8 \\
No SMBH Kinetic Feedback & All SMBH kinetic mode related & See Section 2 in \cite{weinberger.etal.2018} & - \\
Higher SMBH Kinetic Pivot Mass & SMBH kinetic pivot mass ($M_{\rm pivot}$ eqn.~\ref{00}) & $10^{8}M_\odot$ & $4\times10^{8}M_\odot$ \\
Lower SMBH Kinetic Pivot Mass & SMBH kinetic pivot mass ($M_{\rm pivot}$ eqn.~\ref{00})& $10^{8}M_\odot$ & $2.5\times10^{6}M_\odot$ \\
 \end{tabular}}

 \end{center}
 \end{table*}
%%%%%%%%%%%%%%%%%%%%%%%%%%%%%%%%%%%% Table%%%%%%%%%%%%%%%%%%%%%%%%%%%%%
%%%%%%%%%%%%%%%%%%%%%%%%%%%%%%%%%%%%%%%%%%%%%%%%%%%%%%%%%%%
\section{The Effects of stellar and SMBH feedback on SMBH-- hot atmosphere correlations}
\label{sec:theo}
%%%%%%%%%%%%%%%%%%%%%%%%%%%%%%%%%%%%%%%%%%%%%%%%%%%%%%%%%%%%%%%%%%%%%%%
\begin{figure*}
    \centering
 
    \includegraphics[width=0.49\textwidth]{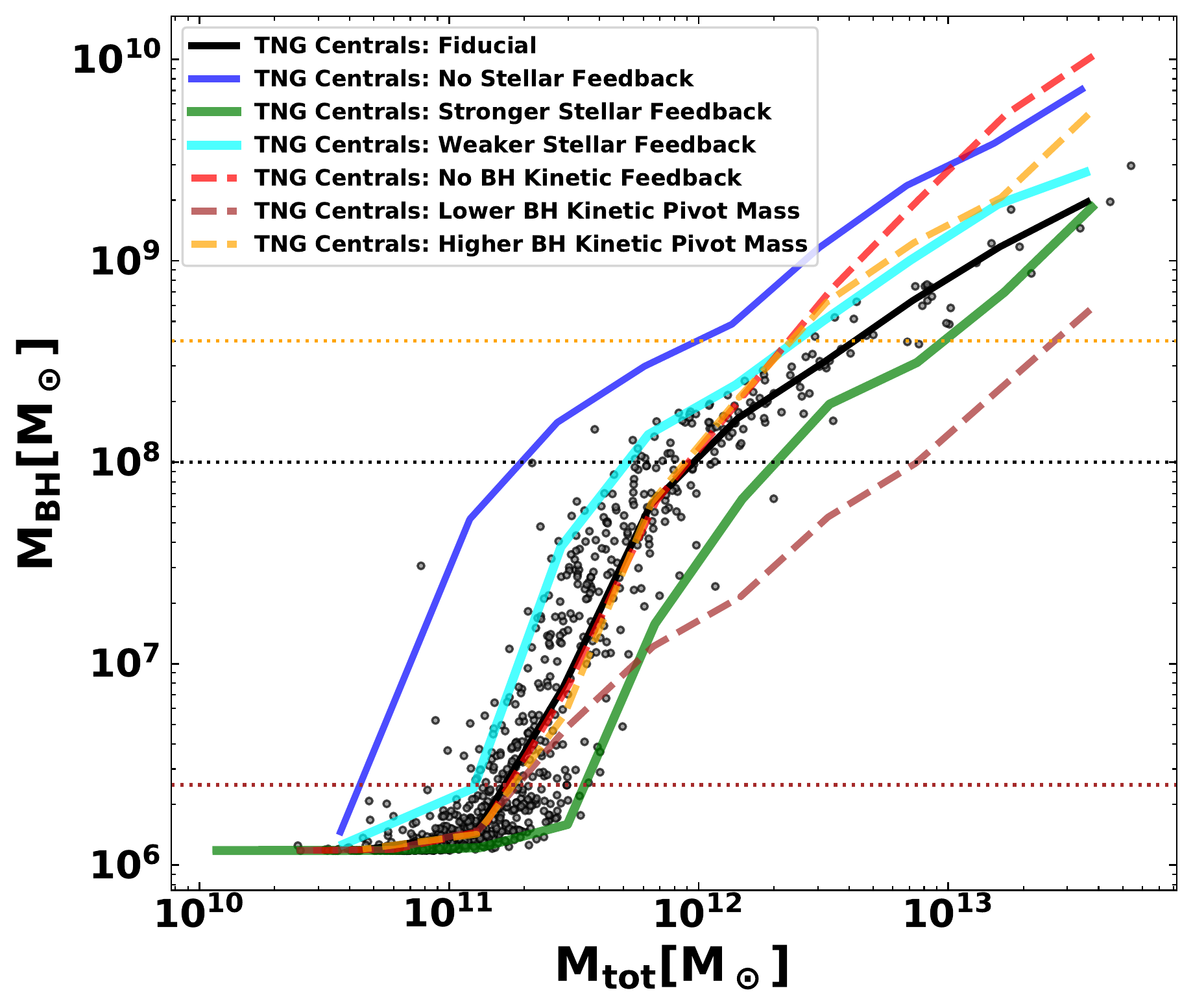}
    \includegraphics[width=0.49\textwidth]{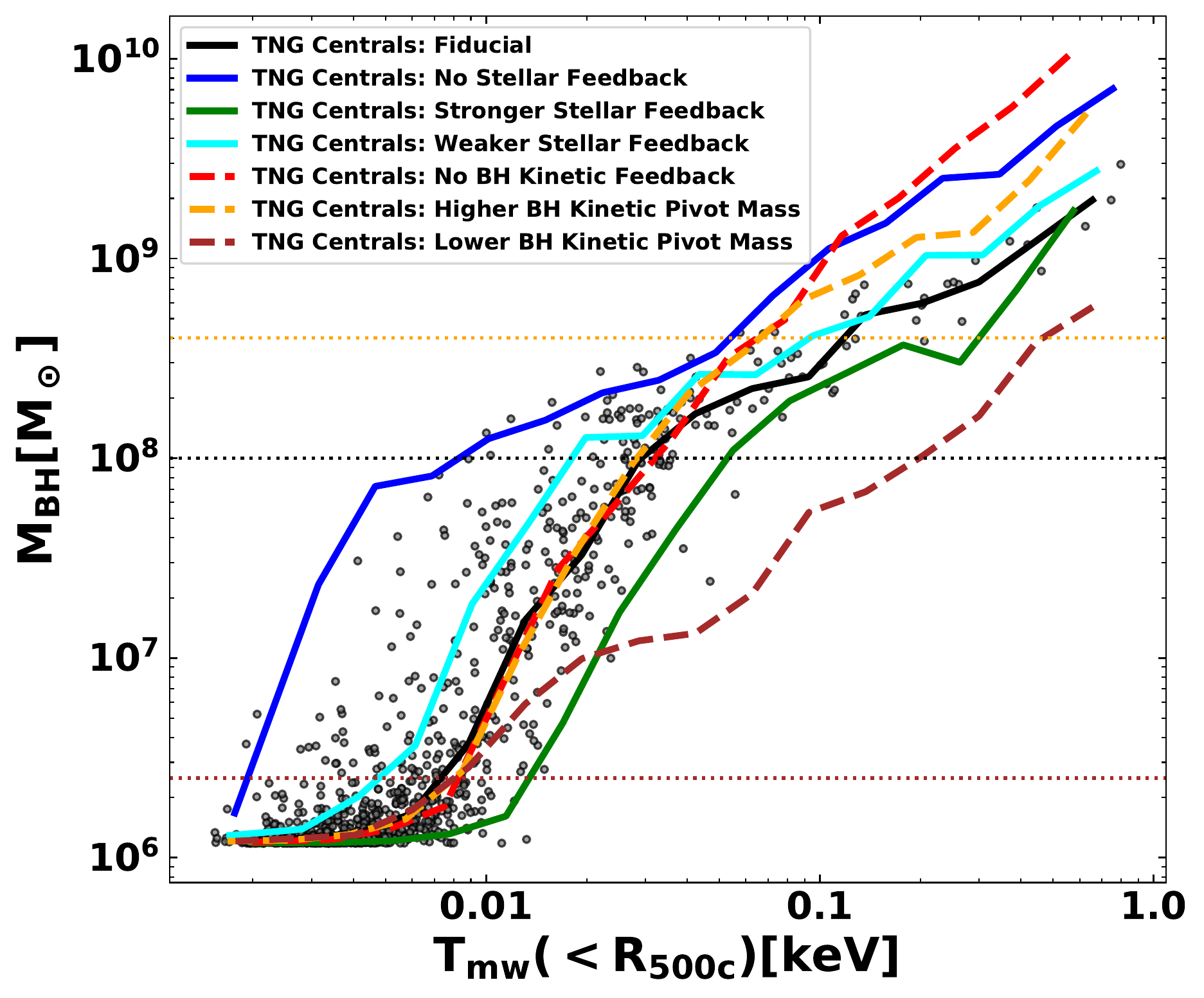}
    \includegraphics[width=0.37\textwidth]{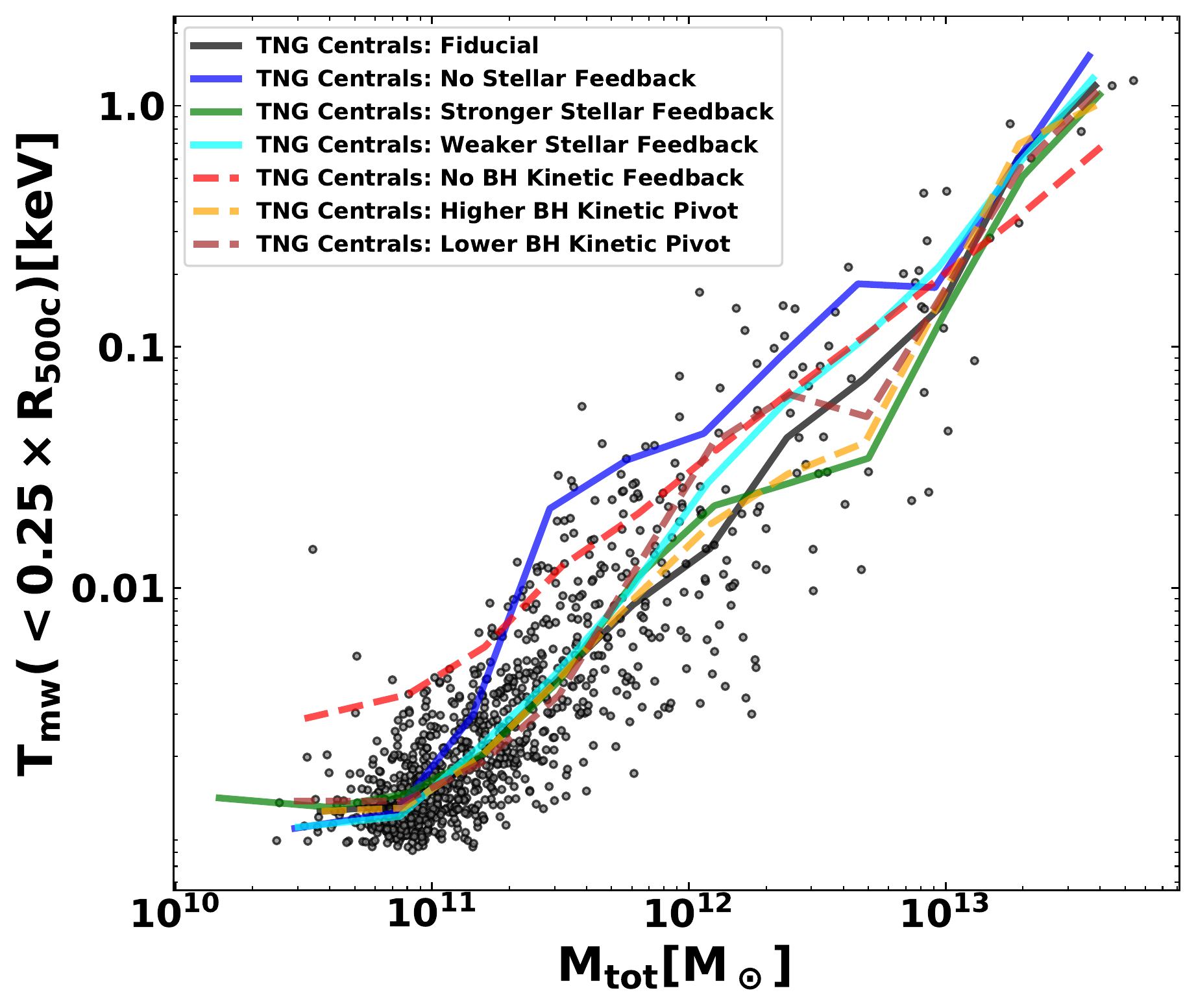}
    \includegraphics[width=0.37\textwidth]{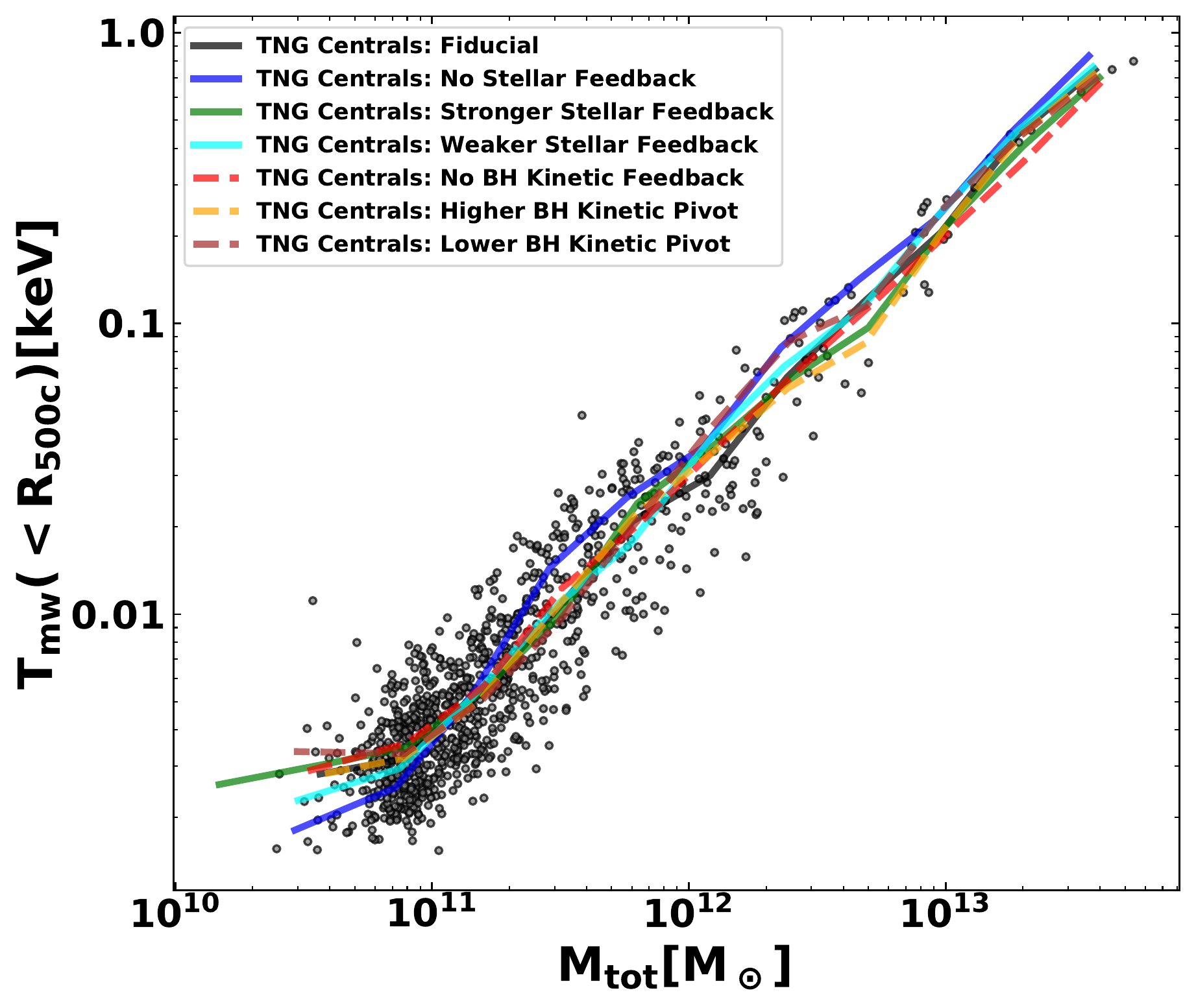}
    \includegraphics[width=0.37\textwidth]{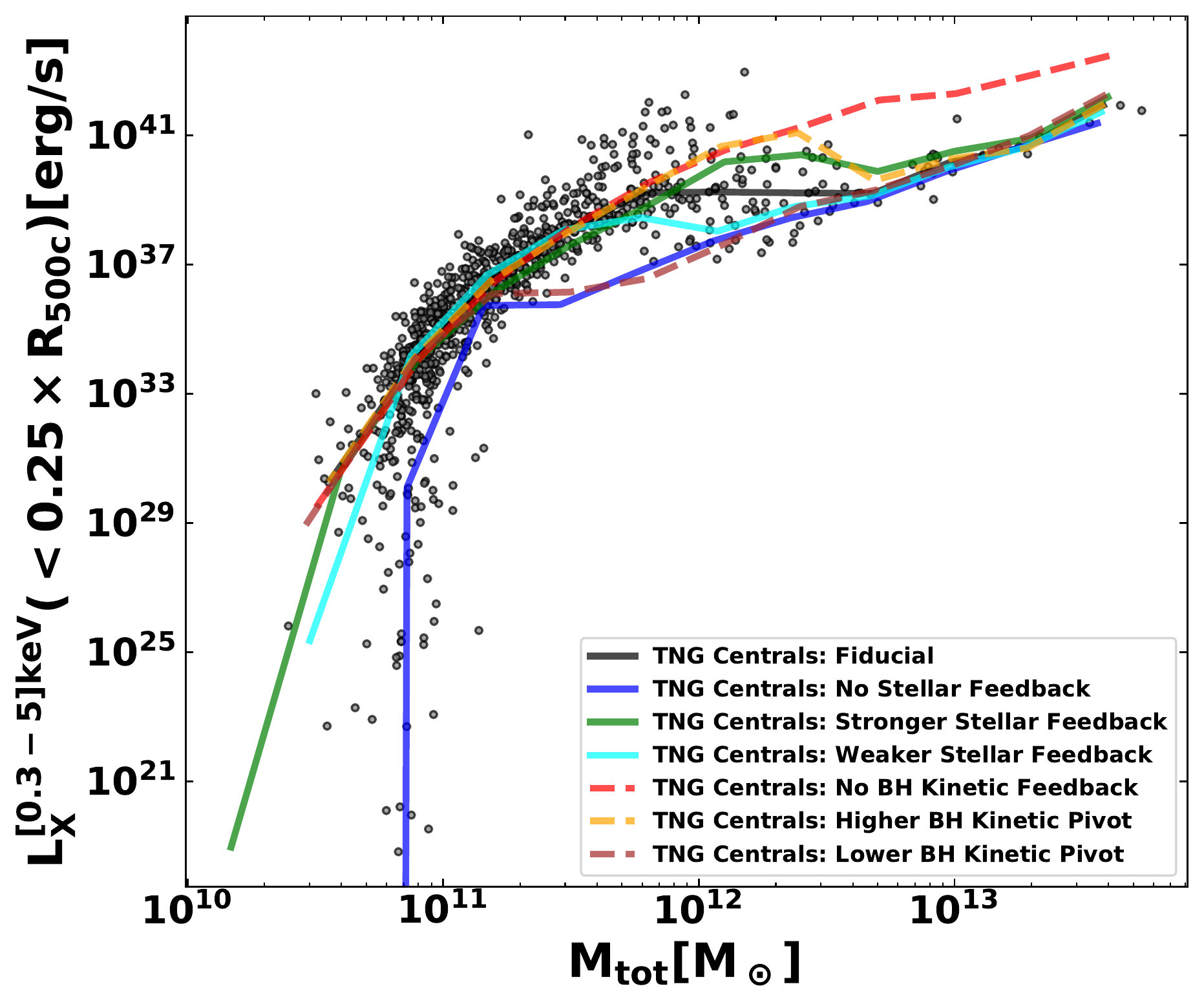}
    \includegraphics[width=0.37\textwidth]{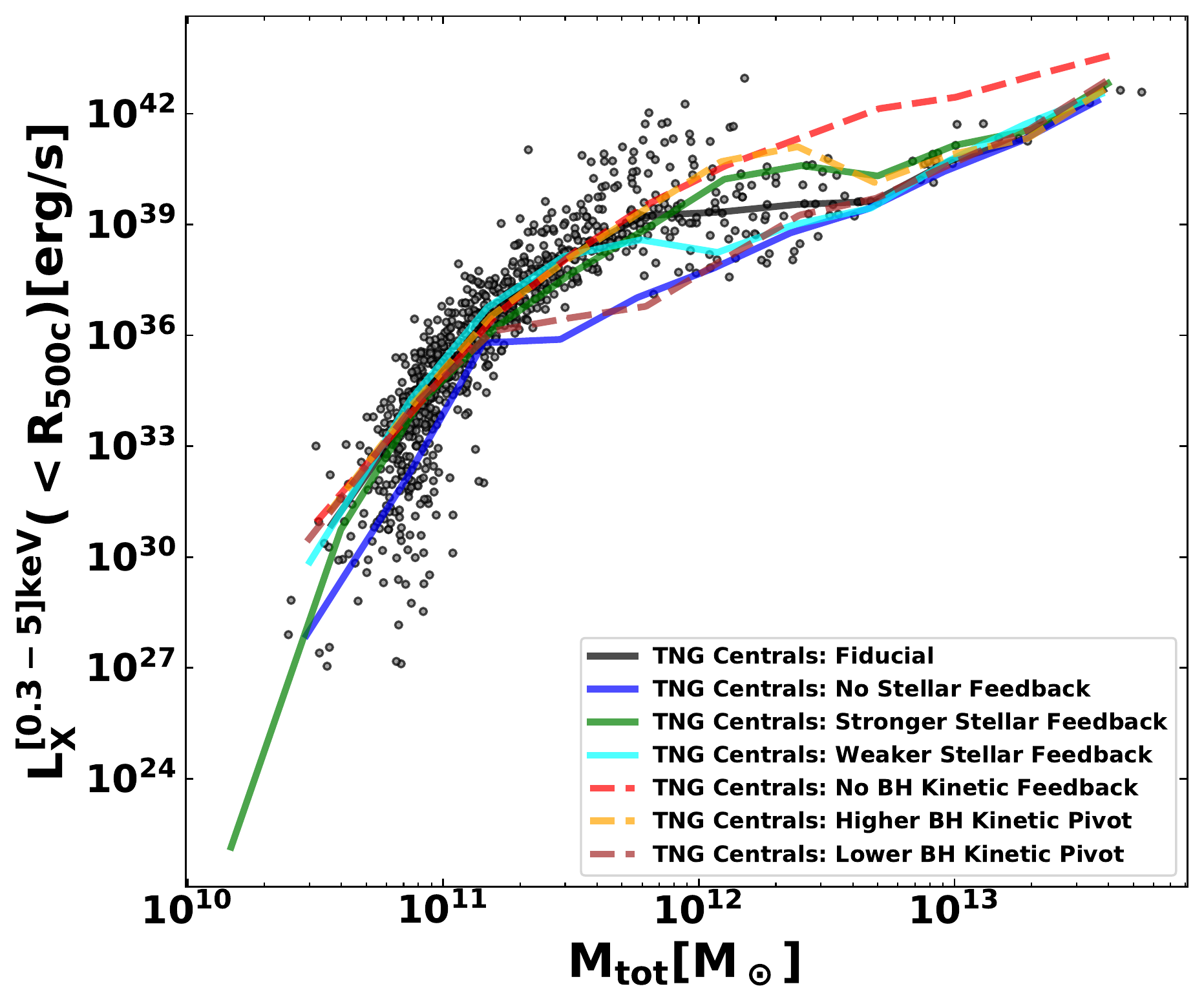}
    \caption{Theoretical investigation on the impact of stellar and SMBH feedback on SMBH-galaxy correlations within the TNG framework. {\it Top left:} The $M_{\rm BH}-M_{\rm tot}$ relations shown for various physical TNG models. {\it Top right:} the corresponding $M_{\rm BH}-T_{\rm mw}$ relations. {\it Four bottom panels: The corresponding $T_{\rm mw}-M_{\rm tot}$ ({\it upper}) and $L_{\rm X}-M_{\rm tot}$ ({\it lower}) relations with the X-ray quantities measured within $0.25\times R_{500c}$ (left) and within $R_{500c}$ (right)}. The solid lines represent the median relations of the models, while data points are obtained from the fiducial TNG model. The horizontal dotted lines mark the pivot SMBH masses at $2.5\times10^{6}M_\odot$, $10^8 M_\odot$, and $4\times10^8 M_\odot$, above which SMBH feedback switches to kinetic mode (see text).}  
    
    \label{fig:4}
\end{figure*}
%%%%%%%%%%%%%%%%%%%%%%%%%%%%%%%%%%%%%%%%%%%%%%%%%%%%%%%%%%%%%%%%%%%%%%%%%%%%%
In the previous Sections, we have shown via the use of the TNG100 simulation that the correlations between SMBHs and the hot gaseous atmospheres are governed by the interplay between gravitational and non-gravitational processes, whose effects manifest differently at different galaxy/halo/SMBH mass scales and at different spatial scales of the considered galaxy or halo gas.  

For high-mass SMBHs at $z=0$ ($M_{\rm BH}\gtrsim 10^8 M_\odot$ or total halo mass $\gtrsim 10^{12} M_\odot$), according to the TNG model, SMBHs and their host haloes can co-evolve by means of the hierarchical assembly, i.e. through mergers (\citealt{weinberger.etal.2018}). On top of that, gas accretion provides an additional channel for the SMBH growth, which dominates at lower SMBH mass scales (or earlier epochs), and which in turn triggers feedback that can further modify the thermodynamics of hot gaseous atmospheres. In addition, evidence has been proposed according to which also feedback driven by star formation can impact the overall SMBH growth (e.g. \citealt{bower.etal.2017}). To illustrate the important role played by non-gravitational physics, in this Section we carry out a theoretical investigation of the effects of stellar and SMBH feedback on the SMBH-hot atmosphere relationships. 

\subsection{Effects of TNG model variations}
For this exercise, we employ a subset of the TNG simulations that were performed on smaller simulation boxes with the size of $\sim25\ \rm{Mpc/h}$ ($\sim37\ {\rm Mpc}$) and resolution very similar to TNG100: these are dubbed L25n512TNG and have been run with various perturbations of the fiducial TNG model. In particular, we focus the exercise on two classes of models: i) models with various strengths of stellar feedback, and ii) models with and without SMBH kinetic feedback and with the latter present at different mass scales. The description of the models are summarised in Table~\ref{tb2}. For a complete exploration of the parameter space of stellar and SMBH feedback models in TNG, the interested reader can refer to \cite{weinberger.etal.2017} and \cite{pillepich.etal.2018}. 

In the top panels of Fig.~\ref{fig:4} we show {two main SMBH} relations for the considered models: $M_{\rm BH}-M_{\rm tot}$ ({\it top left}) and $M_{\rm BH}-T_{\rm mw}$ ({\it top right}) by focusing on the effective temperature of the gas all the way to the halo virial radius, i.e. within $R_{\rm 500c}$.  Additionally, we also present the results on how gaseous properties, i.e. $T_{\rm mw}$ and $L_{\rm X}$, vary among the considered models at various radii: $T_{\rm mw}-M_{\rm tot}$ (middle) and $L_{\rm X}-M_{\rm tot}$ ({\it bottom}) relations with the gaseous properties measured within $0.25\times R_{500c}$ ({\it left}) and within $R_{500c}$ ({\it right}). All the results are shown for all central galaxies in the simulations at $z=0$ with non-vanishing SMBH mass and mass-weighted temperature. For this exercise, we use mass-weighted temperatures to characterise the hot atmospheres: while these are not directly observable, they rather reflect more directly the changes on the state of gas imparted by the different feedback models.

As shown in the top left panel of Fig.~\ref{fig:4} (see also a similar plot on the top right panel of Fig.8 in \citealt[][]{pillepich.etal.2018a}), stellar and SMBH kinetic feedback do have significant and distinct impacts on the final $z=0$ SMBH-halo mass relation, and thus on the growth of SMBHs as a function of halo mass across cosmic epochs. The stellar feedback influences the SMBH growth at the low-mass regime at $z=0$, or equivalently at the early times, with the effect that its presence prevents SMBHs from growing for galaxies with total halo mass $M_{\rm tot}\lesssim10^{11}M_\odot$. Without stellar feedback, low-mass galaxies can enjoy a more efficient SMBH growth even before they reach the total halo mass of about $10^{11}M_\odot$, in comparison to the fiducial TNG model. Above this mass scale, the SMBH growth in the two models with and without stellar feedback is quite similar except that the latter has an overall higher normalisation inherited from its previous growth. For instance, at the high-mass end ($M_{\rm tot}\gtrsim10^{13}M_\odot$), SMBHs in the model without stellar feedback are more massive by about a factor of 3 compared to the fiducial model. Increasing the strength of stellar feedback would delay the fast-growing period of SMBHs to larger mass scales thereby decreasing the overall normalisation of the $M_{\rm BH}-M_{\rm tot}$ relation.  

On the other hand, SMBH feedback in the kinetic mode primarily affects the rate of SMBH growth at the high-mass end, more precisely at the mass scales where the kinetic mode is switched on, i.e. at $M_{\rm BH}\sim2.6\times10^{6},\ 10^{8.2},\ 4\times10^{8} M_\odot$ for the three considered models. Without SMBH kinetic feedback, SMBHs at the high-mass end can experience a faster growth, as evidenced by the steeper slope of the $M_{\rm BH}-M_{\rm tot}$ relation. This result could be explained by the ejective and preventative effect in massive galaxies in TNG, where the SMBH kinetic feedback is efficient enough to prevent gas accretion onto the central SMBH (\citealt{davies.etal.2019,terrazas.etal.2019,zinger.etal.2020}) thereby truncating its otherwise fast growing period driven mainly by gas accretion. By varying the pivotal SMBH mass, which characterises the switch-on threshold of the kinetic mode as given in Eq.~\ref{00}, one can vary the mass range from which the SMBH growth starts to decline.  

When replacing the total halo mass with the mass-weighted temperature within $R_{\rm 500c}$, the $M_{\rm BH}-T_{\rm mw}$ relations, as shown in the top right plot of Fig.~\ref{fig:4}, reflect faithfully the $M_{\rm BH}-M_{\rm tot}$ relations on the left, and preserves the characteristic features of each considered model as discussed above. This result is a direct consequence of the fact that the mass-weighted temperature of the whole circum-galactic or intra-group/cluster gas, $T_{\rm mw}\ (< R_{\rm 500c})$ is in practice a very good proxy of total halo mass, as it appears to be quite insensitive to the considered variations in the underlying simulation physics. This is explicitly shown in the middle-right plot of Fig.~\ref{fig:4}. The $T_{\rm mw}-M_{\rm tot}$ relation {across halo scales} varies insignificantly among the TNG model variations and is statistically consistent with the fiducial TNG model. On the other hand, within smaller volumes of the halo gas and for gas closer to the galaxies' centers ($<0.25R_{500c}$, middle-left plot of Fig.~\ref{fig:4}), the considered model variations do produce $T_{\rm mw}-M_{\rm tot}$ relations that differ from the one obtained from the fiducial model, and by non-negligible amounts, e.g. in comparison to the intrinsic scatter of the fiducial relation. In addition, we notice that the intrinsic scatter of the relations within a smaller aperture are larger than the results within the larger one. For instance, for the fiducial model, the intrinsic scatters of the two $T_{\rm mw}-M_{\rm tot}$ relations at large ($<R_{500c}$) and small ($<0.25R_{500c}$) radii are $0.17$ and $0.24$ dex, respectively. These results imply that the gas temperature at small galactocentric distances is more sensitive to the adopted feedback recipes than when temperatures are averaged throughout halo scales (see also previous TNG analyses, e.g., \citealt{truong.etal.2020,zinger.etal.2020}).

The TNG-inspired findings of the top panels in Fig.~\ref{fig:4} can have far-reaching implications, but also conceal a number of conceptual complexities. Firstly, these results suggest that, even if observations reveal tight correlations between SMBH masses and the X-ray temperature of the gas on group/cluster scales in the present Universe, these correlations do not necessarily need to be a reflection of the direct impact of SMBH feedback on the temperature of the intra-group/cluster gas; rather, they can be an indirect consequence of the way SMBH feedback (and also stellar feedback) determine the availability of gas in the innermost regions of galaxies, that in turn can participate to SMBH growth via accretion, particularly at earlier epochs. 

On the other hand, it is important to mention that when using other estimators of the gaseous temperature such as the emission-weighted $T_{\rm ew}$ (not shown for brevity), the $M_{\rm BH}-T_{\rm ew}$ reflection of the $M_{\rm BH}-M_{\rm tot}$ relation is less perfect than the case of $T_{\rm mw}$: this is due to the fact that the estimation of $T_{\rm ew}$ involves the hot gas content, which instead does greatly depend on the used model. We expand on this last notion in the bottom row of Fig.~\ref{fig:4}, where we show that the  X-ray luminosity of the hot gas, both within halo scales ($<R_{\rm 500c}$) and within smaller galactocentric distances ($<0.25\times R_{\rm 500c}$), is greatly impacted by different implementations of the feedback physics. Among the considered recipes, the model without SMBH kinetic feedback appears to produce the most visibly different $L_{\rm X}-M_{\rm tot}$ relation in comparison to the fiducial outcome. The difference is present even at the high-mass end ($M_{\rm tot}\gtrsim10^{13}M_\odot$), where the other model variations' relations appear to converge toward the fiducial one. For instance, at $M_{\rm tot}\sim10^{13}M_\odot$, the $L_{\rm X}(<R_{500c})$ obtained from the model without SMBH kinetic feedback is larger than the fiducial value by about one order of magnitude. This result can be explained by the ejective effect of SMBH kinetic feedback in TNG model which, as well documented in previous works (e.g. \citealt{weinberger.etal.2018,davies.etal.2019,terrazas.etal.2019,zinger.etal.2020}), reduces the gas density of the galaxy and halo gas, resulting in suppressed gaseous X-ray luminosities.  

\subsection{Summary of the impact of different feedback recipes and speculations on other models}
In summary, the above exercise reveals that, at least within the TNG framework, stellar feedback mainly affects the overall normalisation of the SMBH--hot atmosphere relations, while leaving the shape of the relations largely unaffected\footnote{Note that the variation denoted here as ``No Stellar feedback'' is somewhat different from that denoted as ``No galactic winds'' in \citealt{pillepich.etal.2018a}, where also metal-line cooling is switched off. A comparison between the two reminds us that also the choices on gas cooling are relevant for the determination of SMBH growth.}. On the other hand, SMBH kinetic feedback impacts the SMBH growth rate at the high-mass end in the direction that it truncates the otherwise fast growth enabled by gas accretion. This exercise illustrates how the exact form of the SMBH--hot atmosphere relations depends crucially on both stellar and SMBH feedback, at all spatial and mass scales. On these lines, the interested reader can refer to e.g. \citealt{habouzit.etal.2020} for a detailed comparison of the fiducial Illustris and TNG models on the SMBH-galaxy mass relations.

Finally, at least within the TNG modeling, SMBH feedback does leave an imprint on the thermodynamical (and thus X-ray observable) properties of the circum-galactic and halo gas also in the outer reaches of the gaseous haloes, as theoretically advocated and demonstrated also by, among others, \citealt{gaspari.etal.2019, zinger.etal.2020} and previously e.g. \citealt{barnes.etal.2017, lebrun.etal.2017, truong.etal.2018}. However, within the TNG framework, such imprints are stronger and more direct in the case of the gas density and the X-ray luminosity rather than gas temperature: hence, the observed correlations between the masses of SMBHs and the X-ray temperature of the gas atmospheres in massive (mostly quenched) galaxies ($M_{\rm BH}\gtrsim10^{8.2}M_\odot$) to zeroth order may encode more information about SMBH growth rather than SMBH feedback. 

Our TNG-based results are a direct consequence of the hierarchical assembly of structures coupled with galaxy-scale astrophysical processes, which -- as is the case for most galaxy observables -- prevail in different measures at different galaxy/halo mass scales. For massive SMBHs in massive galaxies and haloes ($M_{\rm BH}\gtrsim10^{8.2}M_\odot$, i.e. $M_{\rm tot}\gtrsim10^{12.5}M_\odot$ ), according to the TNG model it is gravity that mostly determines the growth and hence mass of SMBHs (via mergers) and the thermodynamics of the hot gaseous atmospheres (as predicted by the self-similar ansatz).

We expect this picture to qualitatively hold also in other numerical models, at least those that satisfy two fundamental ingredients and outcomes: i) the explicit inclusion of the hierarchical growth of structures -- no idealized, non-cosmological simulation of isolated systems can make predictions on the nature of SMBH growth if SMBH-SMBH mergers are not included; ii) the return of a population of high mass galaxies with little gas available for SMBH growth via gas accretion -- this being concurrent in TNG with the realization of a population of massive quenched galaxies. However, the exact lower limit of such a high-mass regime can be model dependent. In the TNG framework, the transitional mass scale depends on the details of the two-mode model for SMBH feedback as well as on its interaction with, and the implementation of, the stellar feedback and cooling (see Fig.~\ref{fig:4}); it is also closely related to the characteristic transition mass scale between thermal and kinetic SMBH feedback modes (i.e. $M_{\rm BH}\sim10^8 M_\odot$ in the TNG fiducial model, see Eq.~\ref{00}). However, very importantly, the locus of such characteristic transition mass scale cannot be random, as it is closely related to the stellar mass scale at which galaxies in the Universe transition from mostly star forming to mostly quenched, at least at low redshift (e.g. Figs.~9-10 of \citealt{donnari.etal.2020b}). Therefore, we speculate that models that return very similar galaxy populations in terms of their gas content and star formation activity should correspondingly also return similar SMBH populations and halo gas properties.

%%%%%%%%%%%%%%%%%%%%%%%%%%%%%%%%%%%%%%%%%%%%%%%%%%%%%%%%
\section{Summary and Conclusions}
\label{sec:conclusion}
%%%%%%%%%%%%%%%%%%%%%%%%%%%%%%%%%%%%%%%%%%%%%%%%%%%%%%%%%%%%%%%%%%%%%%%%%%%%%
The recent discovery of tight correlations between the mass of the central SMBH and the X-ray temperature and luminosity of the hot atmosphere permeating the host galaxy suggests that the gaseous component can play an important role in tracing the growth of SMBHs in massive galaxies (\citealt{bogdan.etal.2018,phipps.etal.2019,lakhchaura.etal.2019,gaspari.etal.2019,martin-navarro.etal.2020}). 

In this work, we have investigated the relationship between SMBHs and X-ray properties of hot atmospheres from galactic to cluster mass and spatial scales. In particular, we have employed simulated galaxies from the IllustrisTNG simulations, with a focus on the outcome of the TNG100 run and of a number of simulations where the underlying galaxy physics model was varied. We have carried out a mock X-ray analysis of the simulated galaxies in order to directly compare with X-ray observations. We have discussed the growth of SMBHs in the TNG simulations, their correlation with the properties of the X-ray emitting gas permeating star-forming and quenched galaxies, and the radial dependence within haloes of such correlations, i.e. by considering the hot volume-filling gas within the galactic bodies ($<R_{\rm e}$), within the circum-galactic coronae ($<5R_{\rm e}$), and throughout the intragroup/cluster media ($<R_{\rm 500c}$). We have discussed the simulation outcome within the context of the self-similar model and contrasted it to analytical scaling relations between SMBH masses and gas X-ray temperature and luminosity obtained assuming that SMBH mass traces halo mass. In addition, we have also examined the impact of stellar and SMBH feedback on the shape and scatter of the correlation of the SMBH mass with the properties of the hot atmospheres. Here, we summarise our main results and insights:
\begin{enumerate}

    \item In the TNG framework, the growth of SMBHs can be approximately subdivided into three distinct regimes (Fig.~\ref{fig:1}): a slow- or non-growing phase in low-mass galaxies (with total halo mass $M_{\rm tot}\lesssim10^{11}M_\odot$) also due to stellar feedback, which prevents the accumulation of large amounts of dense gas in the centre of galaxies; a fast-growing period in intermediate-mass galaxies ($10^{11}\lesssim M_{\rm tot}\lesssim10^{12}M_\odot$) via gas accretion; and a declining phase in massive galaxies ($M_{\rm tot} \gtrsim10^{12}M_\odot$), where SMBH feedback heats and expels gas from the centre of galaxies and where SMBHs mostly grow via mergers. We find a strong ($r_{\rm P}\simeq0.87$) and tight ($\sigma_{M_{\rm BH}}\simeq 0.14$ dex) correlation between SMBH masses and their host halo masses in the third regime, which could be a natural outcome of the hierarchical assembly process. We show that, with a typical exposure time of 100 ks with the {\it Chandra X-ray Observatory}, X-ray data can be used to probe parts of the second and third regimes ($M_{\rm tot}\gtrsim10^{11.5}$).  \\
    
    \item Concerning the correlations between the masses of SMBHs and the X-ray properties of the gas within and around galaxies, TNG predicts there exist significant correlations ($|r_{{\rm P}}|>0.5$) in the quenched population (Fig.~\ref{fig:2}). Among the two considered X-ray quantities, the X-ray temperature ($T_{\rm X}$) correlates with SMBH mass better than X-ray luminosity ($L_{\rm X}$). In fact, according to the TNG model, also star-forming galaxies exhibit statistically significant correlations between the mass of SMBHs and the X-ray properties of the gas, but such correlations are weaker ($|r_{\rm P}|\sim0.4-0.5$) than in the case of quenched galaxies.\\
    
    \item According to the TNG model, the correlations of the central SMBH masses with the hot gas properties, $M_{\rm BH}-T_{\rm X}$ and $M_{\rm BH}-L_{\rm X}$, for the quenched population ($M_{\rm BH}\gtrsim10^{8.2}M_\odot$), become more significant and exhibit smaller intrinsic scatters when the effective gas properties are integrated within progressively larger galactocentric distances (Fig.~\ref{fig:e4}). Importantly, the $M_{\rm BH}-T_{\rm X}$ relation at $r\gtrsim5R_{\rm e}$ has a small intrinsic scatter of $\sim0.2$ dex. We also show that the hot gas properties are better connected with total halo mass than with the central SMBH mass, at all spatial scales (Fig.~\ref{fig:e5}). In fact, such correlations between gas properties and total halo mass are in place also in the absence of SMBHs and their feedback, and also for the gas within a few effective stellar radii (Figs.~\ref{fig:e4}, \ref{fig:4}). All these results support a picture whereby the SMBH-hot gas relations are fundamentally a reflection of the $M_{\rm BH}-M_{\rm tot}$ relation.\\

    \item The TNG simulations reproduce reasonably well the observed $M_{\rm BH}-T_{\rm X}$ and $M_{\rm BH}-L_{\rm X}$ relations across the considered mass range, subject to uncertainties in both SMBH mass and X-ray measurements in both simulations and observations (Fig.~\ref{fig:3}). The level of agreement, however, varies among the considered apertures and is the best at the largest radii (within $R_{\rm 500c}$, i.e. at halo scales).\\
    
    %However, the observed relations appear to prefer steeper relations compared to the simulated ones (up to a factor of 2, Fig.~\ref{fig:3}). This result suggests that, while acknowledging the incompleteness of the current X-ray observed samples, simulations predict lower rate of SMBH growth in massive galaxies than expected by the observed data.
    \item On the other hand, we show that the exact form of the SMBH-X-ray relations is sensitive to stellar and AGN feedback, which affect the relations in distinctive ways (Fig.~\ref{fig:4}). For the SMBH mass-temperature relation, particularly for gas on halo scales, this modulation is due to the indirect effects of feedback on the SMBH growth -- via the determination of the availability of gas for accretion --, rather than to a direct effect on the gas temperature. Stellar feedback mainly impacts the relations in the low-mass regimes (i.e. at early times) of the SMBH's growth, suppressing black hole growth and thereby affecting the overall normalization of the relations. SMBH (kinetic) feedback, on the other hand, acts on the SMBH growth in the high-mass (late-time) regimes, by limiting gas availability and by ushering a regime where SMBH growth is dominated by mergers. \\ 
\end{enumerate}

In conclusion, in this paper we have furthered our understanding of the SMBH -- gaseous halo connection suggested by recent X-ray observations and revealed by the existence of tight $M_{\rm BH}-T_{\rm X}$ and $M_{\rm BH}-L_{\rm X}$ correlations. We have demonstrated that the outcome of the TNG simulations is broadly consistent with such observations, lending credibility to the set of insights provided by our analysis. Going beyond what has been accessed observationally so far, the TNG simulations indicate that the SMBH mass should better correlate with the X-ray measurements of the atmospheric temperatures within larger, rather than smaller, galactocentric apertures, in line with the expectations from the self-similar ansatz.  Furthermore, the TNG simulations support a picture whereby such correlations are stronger for quenched (and higher-mass) than for star-forming galaxies. Crucially, within the TNG framework, our analysis suggests that the observed correlations between SMBH and gas properties are primarily inherited from the relationships between SMBH mass and total halo mass ($M_{\rm BH}-M_{\rm tot}$), which at the high-mass end (and only then) are the result of the merging of both SMBHs and dark-matter haloes. In other words, the {\it existence} of the observed correlations between the SMBH mass and the atmospheric gas X-ray temperature and luminosity in TNG is fundamentally a reflection of the underlying relations among SMBH masses, atmospheric gas properties, and total halo mass. In fact, hot gaseous atmospheres develop within haloes in first place not because of the effects of feedback, but because of gravitational collapse within a hierarchical growth of structure scenario. On the other hand, the exact {\it shape and scatter} of the relations do sensitively depend on non-gravitational processes such as star-formation and AGN feedback, so that SMBH feedback can indeed imprint a modulation on such relationships, particularly so at the group and galaxy mass scales and for the atmospheric gas that is closer to the galaxies' or haloes' center.

These results are relevant to the {\it eROSITA} mission (\citealt{merloni.etal.2012}), which is expected to measure the X-ray gas luminosity and temperature for tens of thousands of new clusters, groups, and galaxies. %together with their central AGNs. 
In this regard, our study suggests that the X-ray temperature of the circum-galactic or intra-cluster hot atmospheres measured across halo scales (e.g. within $R_{\rm 500c}$) is the best proxy for the mass of SMBHs at the high-mass end. Conversely, future studies which aim to distinguish between competing feedback scenarios should focus on the X-ray properties of the gaseous haloes in lower-mass objects, e.g. at the group mass scale or below, and should focus on gas properties at smaller radii. Finally, our study indicates that the X-ray luminosity is, at a fixed mass and spatial scale, more sensitive to the SMBH feedback than the X-ray temperature.

%%%%%%%%%%%%%%%%%%%%%%%%%%%%%5
\section*{ACKNOWLEDGEMENTS}
We would like to thank the anonymous referee for constructive feedback and suggestions that helped improve the paper. The authors would like to thank Dylan Nelson for comments on an earlier version of this manuscript and Elad Zinger for useful discussions.
The TNG100 simulation this work is based upon has been run on the HazelHen Cray XC40-system at the High Performance Computing Center Stuttgart as part of project GCS-ILLU of the Gauss centres for Super-computing (GCS, PI: Springel). The TNG model variation runs had been run on the Hydra and Draco supercomputers at the Max Planck Computing and Data Facility (MPCDF, formerly known as RZG) in Garching near Munich. 

%%%%%%%%%%%%%%%%%%%%%%%%%%%%
\section*{Data Availability}
%%%%%%%%%%%%%%%%%%%%%%%%%%%%%%%%%%%%%%%
Data directly related to this publication and its figures are available upon request from the corresponding author. The output of the flagship runs of the IllustrisTNG project is publicly available and accessible at www.tng-project.org/data (\citealt{nelson.etal.19}).
%%%%%%%%%%%%%%%%%%%%%%%%%%%%%%%%%%%%%%%%%%%%%%%%%
%%%%%%%%%%%%%%%%%%%%%%%%%%%%%%%%%%%%%%%%%%%%%%%%%%%%%%%%%%%%%%%%%%%%%%%%%%%%%%5
% Bibliography %
%%%%%%%%%%%%%%%%
\bibliographystyle{mnbst}
\bibliography{ref}
%%%%%%%%%%%%%%%%%%%%%%%%%%%%%%%%%%%%%%%%%%%%%%%%%%%%%%%%%%%%%%%%%%%%%%%%%%%%%%
\appendix
%%%%%%%%%%%%%%%%%%%%%%%%%%%%%%%%%%%%%%%%%%%%%%%%%%%%
\section{Impacts of non-gravitationally bound gas elements}
\label{sec:app1}
%%%%%%%%%%%%%%%%%%%%%%%%%%%%%%%%%%%%%%%%%%%%%%%%%%%%%%%%%%%%%%%%%%%%%%%
%%%%%%%%%%%%%%%%%%%%%%%%%%%%%%%%%%%%%%%%%%%%%%%%%%%%%%%%%%%%%%%%%%%%%%%
\begin{figure*}
    \centering
   \includegraphics[width=0.99\textwidth]{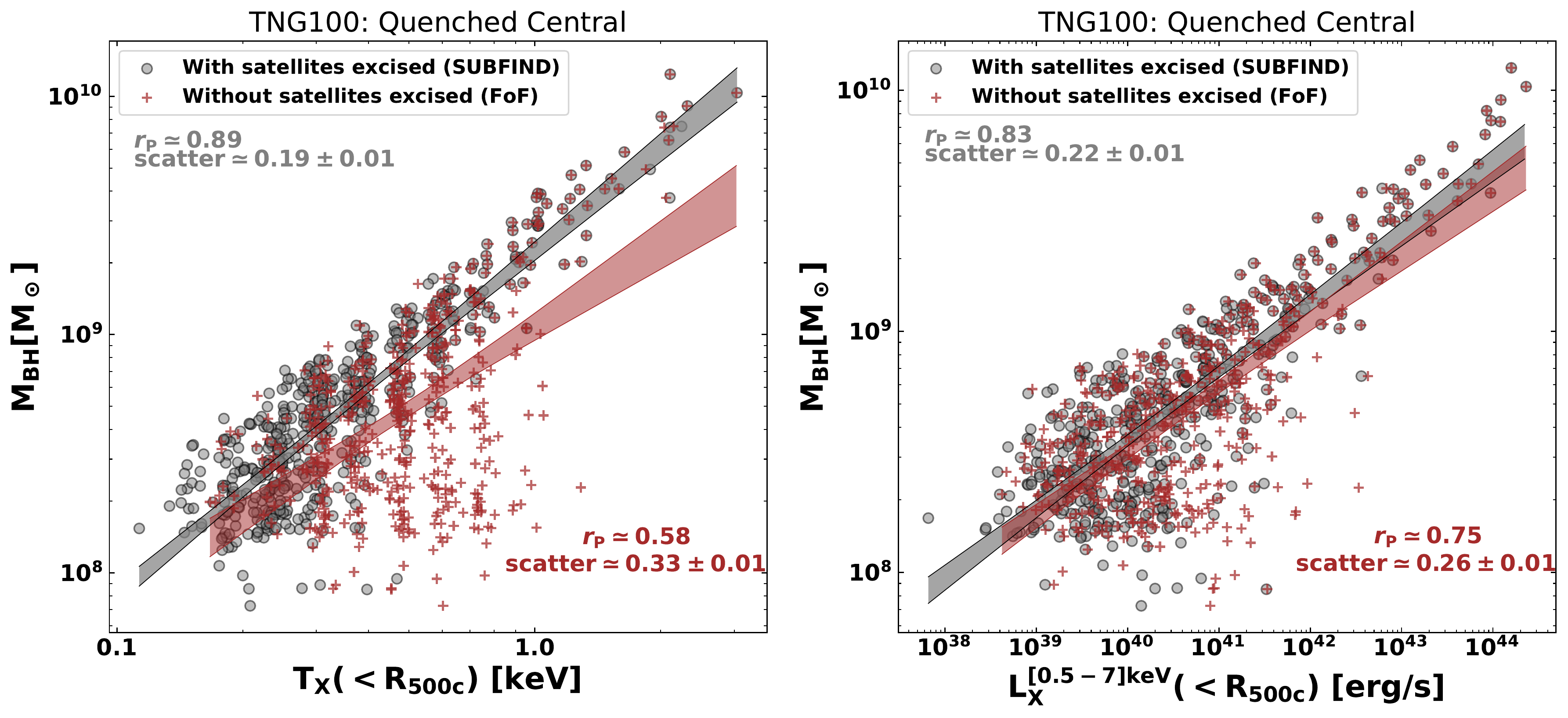}
    \includegraphics[width=0.97\textwidth]{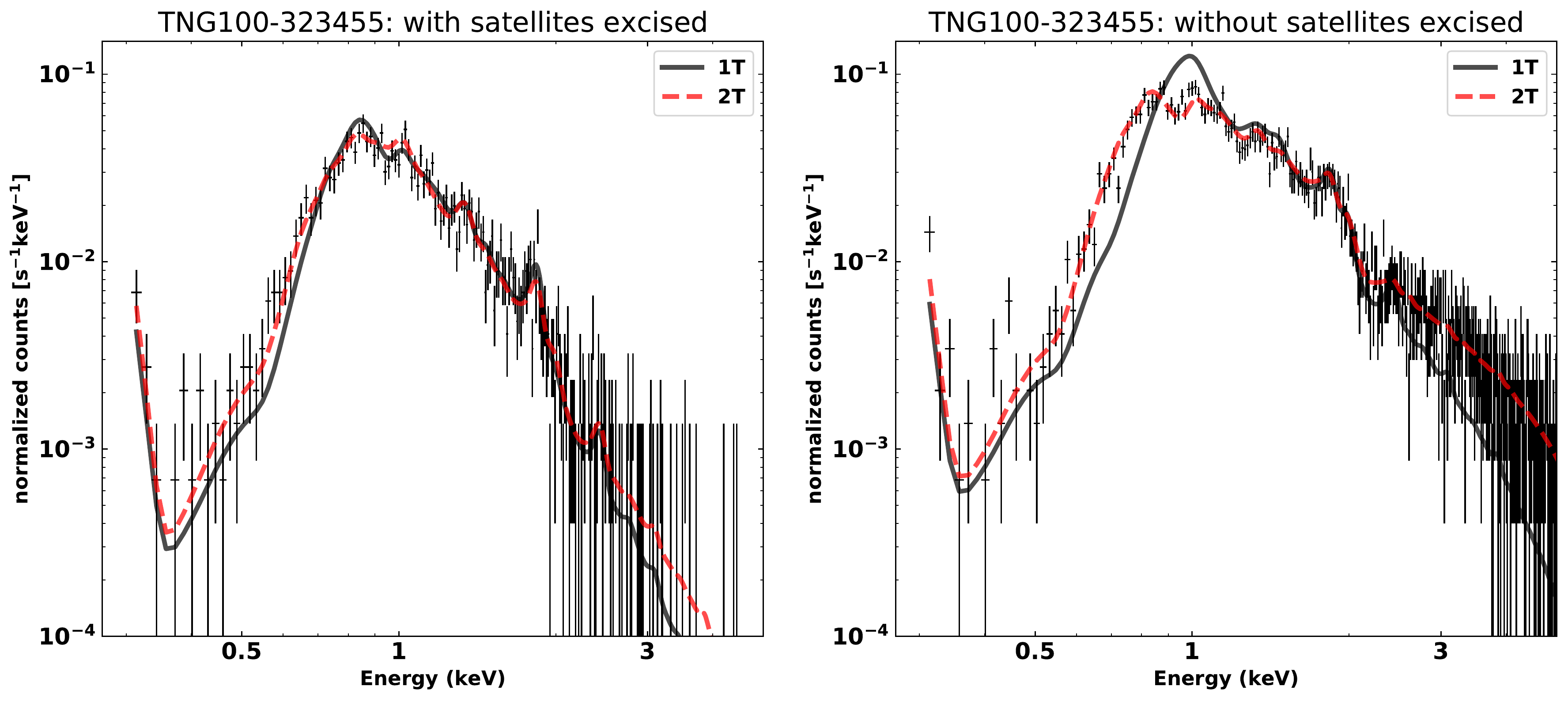}
    \caption{The effects of different gas selections on the SMBH--X-ray scaling relations: with and without satellites excised as implemented in the simulations by adopting {\sc fof} vs. {\sc subfind}-based gas cells. {\it Top row:} The X-ray scaling relations $M_{\rm BH}-T_{\rm X}$ (left) and $M_{\rm BH}-L_{\rm X}$ of intra-cluster medium ($<R_{\rm 500c}$) shown for the two types of gas cells selection. {\it Bottom row:} an example of mock X-ray spectra for a low-mass TNG100 galaxy, for the selection with satellites being excised (left) and without satellites excised (right). The solid and dashed lines are the corresponding best-fit curves for 1T and 2T model, respectively.} 
    \label{fig:a1}
\end{figure*}
%%%%%%%%%%%%%%%%%%%%%%%%%%%%%%%%%%%%%%%%%%%%%%%%%%%%%%%%%%%%%%%%%%%%%%%
We investigate how including non-gravitationally bound gas elements might affect simulated X-ray quantities ($T_{\rm X}$ and $L_{\rm X}$), as well as their relationship with SMBHs mass. To do this, we compare the results obtained from two different types of gas cells selection from simulations: one is based on the {\sc subfind} algorithm, which is the fiducial method we employ in this paper, and the one based on the {\sc fof} algorithm. While the former method selects only gravitationally bound gas elements to the host halo, the latter includes both gravitationally and non-gravitationally bound gas elements, which mainly consist of gas cells attached to subhaloes (satellites) and partly fast-moving elements driven by stellar or SMBH feedback. In simple words, the {\sc subfind} method does excise the contribution of gas cells from satellite galaxies while the {\sc fof} selection does not. In Fig.~\ref{fig:a1}, in the top panels we show the comparison between two types of selection for $M_{\rm}-T_{\rm X}$ and $M_{\rm BH}-L_{\rm X}$ relations of the intra-cluster medium (i.e. measured within $R_{\rm 500c}$). In the bottom panels, we show an example of mock X-ray spectra of a galaxy at the low-mass end, where the two selections display the most pronounced difference.

The main effect of using the {\sc fof} selection, compared to the {\sc subfind} method adopted throughout this paper, is that it causes the X-ray scaling relations to be less significant, with larger scatters. For instance, the {\sc fof}-based $M_{\rm BH}-T_{\rm X}$ relation ($r_{\rm P}\sim0.58$) is much weaker than the {\sc subfind}-based counterpart ($r_{\rm P}\sim0.89$). The former also exhibits remarkably higher intrinsic scatter than the latter, by more than $0.1$ dex. This is due to, as visible in the top-left panel of Fig.~\ref{fig:a1}, the low-temperature systems ($T_{\rm X}\lesssim1$ keV) whose temperature increases significantly in the case of the {\sc fof} gas selection, i.e. when the contribution from satellite gas is not excised. The change in the temperature of those systems also modifies the shape of the $M_{\rm BH}-T_{\rm X}$ relation, especially its normalisation, which in the case of the {\sc fof} selection is about a factor of two lower than the corresponding {\sc subfind}-based value. On the other hand, the best-fit {\sc fof}-based $M_{\rm BH}-L_{\rm X}$ relation varies insignificantly compared to the {\sc subfind}-based, given the relation intrinsic uncertainties. 

 Further inspection of mock X-ray spectra of an example low-mass galaxy, as shown in the bottom row of Fig.~\ref{fig:a1}, reveals that gas phase in the case of {\sc fof} selection accommodates more hot temperature components than in the case of {\sc subfind} selection. This is evident, as shown in the bottom plots, that the {\sc fof}-selected mock spectrum clearly prefers a two-temperature model (2T) over a single-temperature model (1T), while for the {\sc subfind}-selected mock spectrum the two models fit approximately equally well. The hotter components might come from hot atmospheres associated with satellite galaxies that belong to the same halo, or they might be partly of highly-ejective gas cells, which could be of high temperature, driven by stellar or SMBH feedback. When including those extra components into the mock X-ray spectra, the resulting best-fit temperature no longer closely reflects the potential well of the central galaxy, thereby lessening the significance of the SMBH mass--temperature relation. We note again that this effect is mostly prominent in low-temperature (or low-mass) systems, while in massive systems the gravitational potential well predominantly determines the gas temperature. We discuss at the end of Section~\ref{sec:comparison_xray} how this result might be relevant for observational studies of the $M_{\rm BH}-T_{\rm X}$ relation.    
\end{document}